\documentclass[]{aa}
\usepackage{graphicx,txfonts,amssymb,natbib}
\authorrunning{C. Fedeli et al.}
\titlerunning
  {Clustering with dynamical-dark energy}

\begin{document}

\title
  {Observing the clustering properties of galaxy clusters in dynamical dark-energy cosmologies}

\author{C. Fedeli\inst{1,2,3}\thanks{E-mail: cosimo.fedeli@unibo.it}, L. Moscardini\inst{1,3} \and M. Bartelmann\inst{4}}
 
 \institute{$^1$
     Dipartimento di Astronomia, Universit\`a di Bologna,
     Via Ranzani 1, 40127 Bologna, Italy\\$^2$ INAF-Osservatorio
     Astronomico di Bologna, Via Ranzani 1, 40127 Bologna, Italy\\$^3$
     INFN, Sezione di Bologna, viale Berti Pichat 6/2, I-40127 Bologna, Italy\\$^4$
     Zentrum f\"ur Astronomie, ITA, Universit\"at Heidelberg,
     Albert-\"Uberle-Str. 2, 69120 Heidelberg, Germany}
 
\date{\emph{Astronomy \& Astrophysics, in press}}

\abstract{We study the clustering properties of galaxy clusters expected to be observed by various forthcoming surveys both in the X-ray and sub-mm regimes by the thermal Sunyaev-Zel'dovich effect. Several different background cosmological models are assumed, including the concordance $\Lambda$CDM and various cosmologies with dynamical evolution of the dark energy. Particular attention is paid to models with a significant contribution of dark energy at early times which affects the process of structure formation. Past light cone and selection effects in cluster catalogs are carefully modeled by realistic scaling relations between cluster mass and observables and by properly taking into account the selection functions of the different instruments. The results show that early dark-energy models are expected to produce significantly lower values of effective bias and both spatial and angular correlation amplitudes with respect to the standard $\Lambda$CDM model. Among the cluster catalogs studied in this work, it turns out that those based on \emph{eRosita}, \emph{Planck}, and South Pole Telescope observations are the most promising for distinguishing between various dark-energy models.}

%\begin{keywords}
%gravitational lensing --- galaxies: clusters --- cosmology: theory ---
%dark matter 
%\end{keywords}

\maketitle

\section{Introduction}

One of the main quests of contemporary astrophysics is the determination of the nature of dark energy, the mysterious component of the cosmic fluid responsible for the accelerated expansion of the Universe. While the evidence of its presence has became compelling in the past decade \citep{AS06.1,RI07.1,DU08.1,KO08.1,RU08.1,KI08.1}, its nature remains entirely unexplained. In particular, while virtually all the present observational evidence is in concordance with a cosmological-constant interpretation of the dark energy, its possible dynamical evolution is not well constrained. The detection of this time evolution would hint at dark energy being different from vacuum energy, and would call for some more general explanation, such as minimally coupled scalar fields \citep{RA88.1,WE88.2,BR00.1}. Since, in this standard generalization, dark energy is not supposed to clump on the scales of the largest cosmic structures, the only way its nature can be unveiled is by studying the expansion history of the Universe. The expansion rate of the Universe as a function of cosmic time in turn affects the process of structure formation, and consequently the many observable properties of cosmic structures that are accessible to observations. 

The most immediate effect of cosmology on cosmic structures is on the number counts of objects, especially the most massive and extreme ones such as galaxy clusters. A generic dynamical dark-energy model that postulates a dark-energy density increasing with redshift will necessarily imply an earlier structure formation than in cosmological-constant models (if the linear amplitude of density fluctuations at present is fixed), and thus a higher abundance of objects at any time.

An alternative channel for the detection of possible effects of dynamical-dark energy is the study of clustering properties of cosmic structures. Models predicting a higher abundance of objects imply that high-mass clusters are less exceptional, and as a consequence the linear bias with respect to the underlying dark-matter density-field should be reduced. The linear bias, the structure abundance, and the linear correlation function of density fluctuations all affect the angular and spatial correlation functions that are observed in cluster catalogs, and all depend on the behavior of dark energy. We are therefore justified in exploring the effect of different quintessence models on the clustering properties of massive galaxy clusters, and understanding whether differences between them could be detected significantly in cluster catalogs produced by forthcoming experiments. This is the purpose of the present work.

We focus on blind cluster surveys based on the X-ray emission of the hot intra-cluster plasma and on the spectral distortion of the CMB radiation produced by the thermal Sunyaev-Zel'dovich (\citealt{SU72.1}, SZ henceforth) effect. The cosmological models that we address range from the concordance cosmological-constant $\Lambda$CDM cosmogony to early dark-energy models, to intermediate models with a constant equation-of-state parameter for dark energy $w_\mathrm{x} \ne -1$ or with a gentle evolution in $w_\mathrm{x}$ with time. \cite{SA07.1} performed a simple preliminary study of the two-point angular correlation function predicted to be observed by \emph{Planck} in models with early quintessence. As will become evident from the results shown in this paper, our findings are qualitatively consistent with those, while a quantitative comparison is not directly possible because of the different catalog definitions. Observational results about the clustering properties of galaxy clusters measured in optical and infrared catalogs can be found in \cite{BR07.1}, \cite{PA08.1} and \cite{ES08.1}.

This paper is structured as follows. In Sect.~\ref{sct:cosmo}, a brief overview of the different cosmological models employed in this paper is presented, with a description of their main features and a summary of the various cosmological parameters. In Sect.~\ref{sct:clus}, we review the formalism used to describe clustering of galaxy clusters in the past light cone of the observer. In Sect.~\ref{sct:survey}, we detail the properties of the X-ray and SZ surveys analyzed in the present work and summarize the scaling laws used in linking the mass and redshift of objects to their observable properties in Sect.~\ref{sct:scaling}. In Sect.~\ref{sct:cat}, we describe the properties of the cluster catalogs obtained therefrom. In Sect.~\ref{sct:res}, we present our results on the spatial and the angular correlation functions, and in Sect.~\ref{sct:sum}, we summarize our conclusions. We shall use throughout the \cite{SH02.1} prescription for the computation of both the cluster mass function and the linear bias.

\section{Cosmological models}\label{sct:cosmo}

We use seven different cosmological models. The first four of them are models with an early dark-energy (EDE henceforth) component, labeled from EDE1 to EDE4. In early dark-energy cosmologies, the dark-energy contribution is assumed to be represented by a quintessence scalar field whose evolution tracks that of the dominant component of the cosmic fluid at a given time \citep{WE88.1,WE88.2,WE95.1}. As a consequence, the density parameter for dark energy at very early times does not vanish as in more conventional models, but flattens to a finite value. To ensure dark-energy dominance at low redshift however, an \emph{ad hoc} mechanism for breaking the tracking behavior must be adopted, usually in the form of a non-standard kinetic term in the quintessence Lagrangian \citep{HE01.1,DO01.1,DO06.1} or a non-minimal coupling between quintessence and neutrinos with evolving mass \citep{WE07.1}.

An adequate parametrization of early dark-energy models consists of the dark-energy density parameter at present, $\Omega_{\mathrm{x},0}$, the dark-energy equation-of-state parameter at present, $w_{\mathrm{x},0}$ and a suitable average of the dark-energy density parameter at early times during the phase of linear structure formation,
\begin{equation}
\bar{\Omega}_\mathrm{x,sf} \equiv -\frac{1}{\ln a_\mathrm{eq}} \int_{\ln a_\mathrm{eq}}^0 \Omega_\mathrm{x} (a) d\ln a,
\end{equation}      
where $a_\mathrm{eq}$ is the scale factor at matter-radiation equality. Observational constraints from Type Ia supernovae, large-scale structure, and CMB allow $\bar{\Omega}_\mathrm{x,sf}$ to be on the order of a few percent at most \citep{DO05.1,DO07.1}. Our models EDE1 and EDE2 were introduced and studied in \cite{BA06.1} (see also \citealt{FE07.1}), while EDE3 and EDE4 were investigated by \cite{WA08.1} and have cosmological parameters more closely adapted to the latest WMAP data releases \citep{DO06.1,DO07.1}. For a detailed analysis of how early dark-energy models compare with other dark-energy models on observational grounds, especially with respect to type-Ia supernovae data sets, we refer the reader to \cite{RU08.1}.

Apart from the EDE models, we also investigate a cosmological model with dynamical dark-energy parametrised as in \cite{KO08.1}, which we briefly describe below. In this case, the dark-energy equation-of-state parameter is assumed to evolve with the scale factor as
\begin{equation}\label{eqn:wx}
w_\mathrm{x}(a) = \frac{aw_0}{a+a_*} + \frac{a(1-a)w_1}{a+a_*} - \frac{a_*}{a+a_*},
\end{equation}
where $z_* = 1/a_*-1$ is a transition redshift that we set to be $z_* = 10$ in what follows. In any case, \cite{KO08.1} showed that the precise choice of the transition redshift is not extremely relevant to the inferred value of the parameters $w_0$ and $w_1$. In the low-redshift limit, $z \ll z_*$, Eq. (\ref{eqn:wx}) reduces to the more standard form
\begin{equation}
w_\mathrm{x}(a) = w_0 + (1-a)w_1
\end{equation}
\citep{CH01.1,LI03.1}. We shall assume in the following that $w_0 = -1$, while $w_1 = 0.5$, which is almost the highest possible value inferred at $95.4\%$ confidence level by WMAP-5 year data \citep{KO08.1}. For brevity, we refer to this model by K08 in the remainder of the paper.

\begin{figure}[t]
  \includegraphics[width=\hsize]{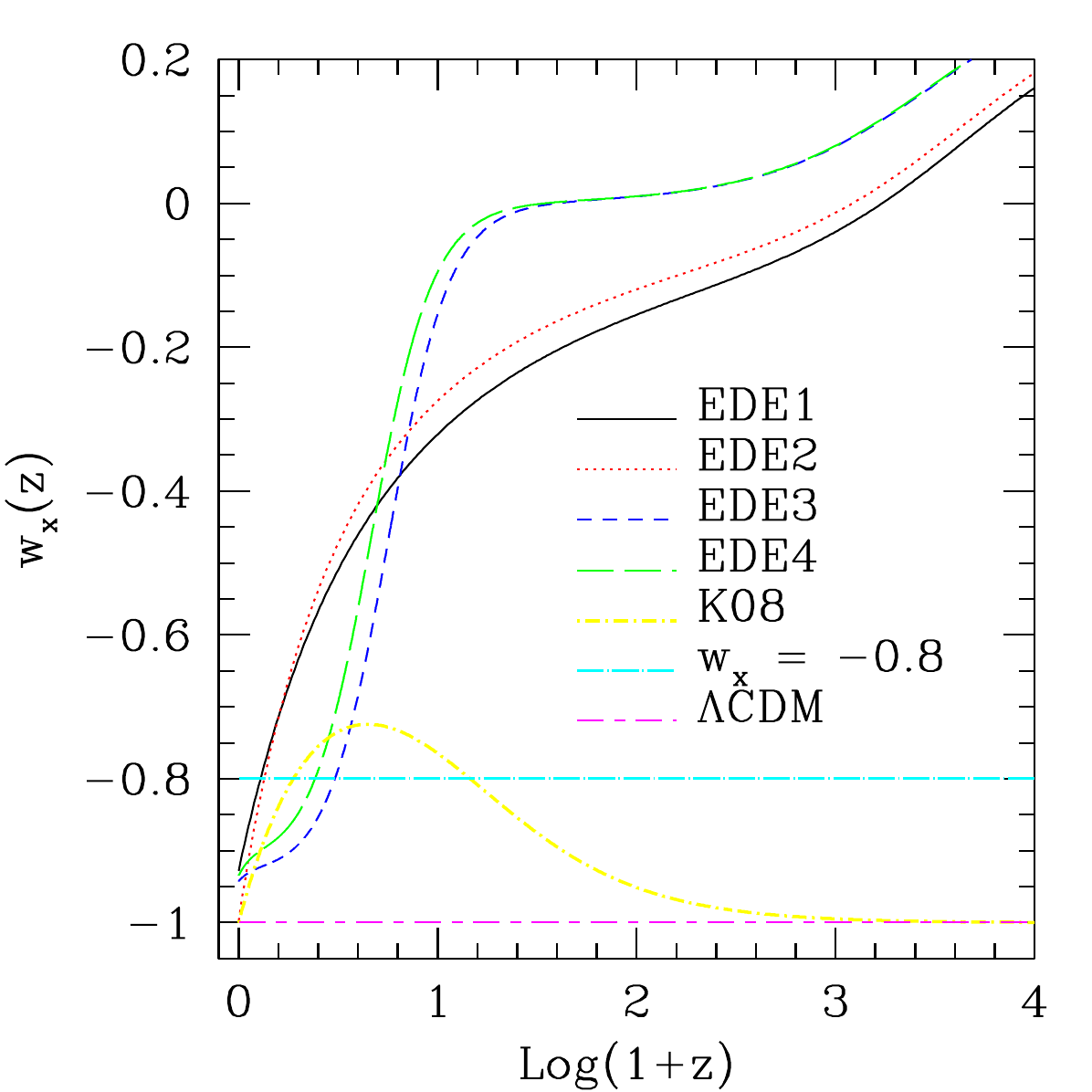}\hfill
\caption{The redshift evolution of the dark energy equation of state parameter $w_\mathrm{x}$ for the seven dark energy models adopted in this work, as labelled in the plot.}
\label{fig:wz}
\end{figure}

In addition to those described above, we analyzed a model with constant $w_\mathrm{x} = -0.8$ and a standard $\Lambda$CDM cosmological model with parameters given by the latest WMAP-5 year data release in combination with Type-Ia supernovae and baryon acoustic oscillations \citep{KO08.1}. The redshift evolution in the equation-of-state parameters for our seven dark-energy models is shown in Fig.~\ref{fig:wz}, while the values of the main cosmological parameters are summarized in Table~\ref{tab:par}. There, the Hubble constant is expressed as $H_0 = h\: 100$ km s$^{-1}$ Mpc$^{-1}$, and $\sigma_8$ represents the \emph{rms} of primordial density fluctuations smoothed on a scale of $8\,h^{-1}$ comoving Mpc.

\begin{table}[t!]
  \caption{Parameter values for the seven cosmological models investigated in this work}
  \begin{center}
    \label{tab:par}
    \begin{tabular}{|c|c|c|c|c|c|c|}
      \hline
      Model & $\Omega_{\mathrm{m},0}$ & $\Omega_{\mathrm{x},0}$ & $\bar{\Omega}_{\mathrm{x,sf}}$ 
      & $h$ & $w_{\mathrm{x},0}$ & 
      $\sigma_8$\\
      \hline
      \hline
      EDE1 & $0.330$ & $0.670$ & $0.040$ & $0.670$ & $-0.928$ & $0.820$ \\
      EDE2 & $0.360$ & $0.640$ & $0.040$ & $0.620$ & $-0.997$ & $0.780$ \\
      EDE3 & $0.284$ & $0.716$ & $0.033$ & $0.686$ & $-0.942$ & $0.715$ \\
      EDE4 & $0.282$ & $0.718$ & $0.048$ & $0.684$ & $-0.935$ & $0.655$ \\
      K08 & $0.279$ & $0.721$ & $-$ & $0.701$ & $-1$ & $0.817$ \\
      $w_\mathrm{x} = -0.8$ & $0.279$ & $0.721$ & $-$ & $0.701$ & $-0.800$ & $0.817$ \\
      $\Lambda$CDM & $0.279$ & $0.721$ & $-$ & $0.701$ & $-1$ & $0.817$ \\
      \hline
    \end{tabular}
  \end{center}
\end{table}

Note the highly non-trivial behavior of $w_\mathrm{x}(z)$ in the early dark-energy models, for which the equation-of-state parameter reaches positive values for relatively low redshift, especially for EDE3 and EDE4. We also note the extremely low normalization of the power spectrum for linear density fluctuations (parametrized by $\sigma_8$) in model EDE4, necessary to counter act the quite high dark-energy density at early times, which in turn determines a particularly low linear density contrast threshold for spherical collapse. As can be seen from Fig.~\ref{fig:wz}, all our models approach the cosmological-constant behavior at low redshift, and they are all constructed to also be in agreement, beyond CMB data, with large-scale structure \citep{TE04.2,TE04.1} and Type-Ia supernovae data.

\section{Clustering formalism}\label{sct:clus}

We used the formalism developed by \cite{MA97.1} and further applied, among others, by \cite{MO98.1} in studying high-redshift galaxy clustering and by \cite{MO00.1,MO01.1,MO02.1} in describing the clustering of galaxy clusters in the past light cone of an observer given a survey selection function. In this section, we briefly summarize this formalism and refer to the quoted papers for additional detail.

The starting point of the formalism is the number of objects of mass $M$ per unit redshift around $z$ given a background cosmology, $\mathcal{N}(M,z) = 4\pi g(z) n(M,z)$, where $n(M,z)$ is the standard differential mass function \citep{PR74.1,BO91.1,SH02.1} and $g(z)$ is the Jacobian determinant
\begin{equation}\label{eqn:jacobian}
g(z) = r^2(z) \frac{dr}{dz}(z)\,.
\end{equation}
Evaluating the differential mass function for cosmological models with dynamical dark-energy, we set the linear-density threshold for collapse as computed in \cite{BA06.1}. We shall further comment on this choice in Sect.~\ref{sct:sum}. In Eq.~(\ref{eqn:jacobian}), $r(z)$ is the comoving radial distance to redshift $z$, and $g(z)$ thus represents the comoving volume per unit redshift around $z$. Equation~(\ref{eqn:jacobian}) is valid only for a spatially flat cosmological model, which we assume henceforth. Under the same assumption, the comoving radial distance $r(z)$ is
\begin{equation}
r(z) = \frac{c}{H_0} \int_0^z \frac{dz'}{E(z')}\,,
\end{equation}
where $E(z) \equiv H(z)/H_0$ is the normalised Hubble parameter.

We consider a galaxy-cluster catalog covering the mass range $\left[ M_1,M_2 \right]$, where $M_1$ and $M_2$ in general depends on redshift. Then, the true all-sky equivalent redshift distribution of objects in the catalog reads
\begin{equation}\label{eqn:dist}
\mathcal{N}(z) = \int_{M_1}^{M_2} \mathcal{N}(M,z) dM.
\end{equation}
For the realistic situations explored below, the observed redshift distribution equals Eq.~(\ref{eqn:dist}) multiplied by the fractional sky coverage $f_\mathrm{sky}$ of the survey. In the real world, cluster catalogs are restricted by a threshold value of some observable (for instance the X-ray flux or the SZ decrement), hence all objects of mass higher than the mass $M_1$, corresponding to this limiting observable at a given redshift by some scaling relation, will be included in the catalog. This amounts to setting $M_2 = + \infty$ in Eq.~(\ref{eqn:dist}).

One important ingredient for predicting the observed clustering properties of galaxy clusters is a relation between the density contrast of collapsed objects and that of the underlying matter distribution with the correct time evolution, i.e.,~the effective bias. For the linear `monochromatic' bias, we adopt the expression
\begin{eqnarray}
b(M,z) &=& 1 + \frac{1}{\delta_\mathrm{c}} \left[ a \frac{\delta_\mathrm{c}^2}{D_+^2(z)S(M)} -1 \right] + 
\nonumber
\\
&+& p\frac{2}{\delta_\mathrm{c}} \left[ \frac{1}{1 + \left[ \sqrt{a}\delta_\mathrm{c}/\left(D_+(z)\sqrt{S(M)} \right) \right]^{2p}} \right],
\end{eqnarray}
where $D_+(z)$ is the linear growth factor for density fluctuations, $S(M)$ is the variance in the primordial density field smoothed on a scale corresponding to the mass $M$, and $\delta_\mathrm{c}$ is the linearly extrapolated density contrast at collapse for a spherical density perturbation. We note that in general $\delta_\mathrm{c}$ depends on redshift (although mildly for models with constant $w_\mathrm{x}$), but for clarity we omitted this dependence in the previous equation.

The standard \cite{PR74.1} (see also \citealt{MO96.1}) relation is recovered by setting $a = 1$ and $p = 0$. The more precise relation proposed by \cite{SH99.1} (see also \citealt{SH01.1}) can be obtained by instead setting $a = 0.707$ and $p = 0.3$. We shall follow this second option for consistency with the mass-function prescription and because it has been shown to provide closer agreement with the bias measured in $\Lambda$CDM numerical simulations.

Given the above, the effective bias is defined as the linear bias weighted by the abundance of clusters in the catalog at hand,
\begin{equation}\label{eqn:beff}
b_\mathrm{eff}(z) \equiv \frac{1}{\mathcal{N}(z)} \int_{M_1}^{M_2} b(M,z) \mathcal{N}(M,z) dM\,.
\end{equation}

We define $\xi(r,z_1,z_2)$ to be the two-point correlation function of the underlying density distribution, for density peaks placed at the two different redshifts $z_1$ and $z_2$. This is conveniently computed by Fourier-transforming the non-linear power spectrum of the density fluctuations, described e.g.,~using the fit of \cite{PE96.1}. More accurate prescriptions for evaluating the non-linear matter power spectrum exist \citep{SM03.1}, but their differences to that given by \cite{PE96.1} are small and relevant only on scales $\lesssim 1$ Mpc $h^{-1}$, which are not pertinent here. The object correlation function can then be defined as 
\begin{equation}
\xi_\mathrm{obj}(r,z_1,z_2) \equiv b_\mathrm{eff}(z_1)b_\mathrm{eff}(z_2)\xi(r,z_1,z_2)\,.
\end{equation}
The problem given by the presence of a double redshift dependence in the correlation function of the underlying density fluctuations is solved by considering a single, average redshift $\overline{z}$ defined as $D_+(\overline{z}) = \sqrt{D_+(z_1)D_+(z_2)}$ so that, effectively, $\xi(r,z_1,z_2) = \xi(r,\overline{z})$. The effect of redshift-space distortions is also taken into account by multiplying the correlation function $\xi(r,z_1,z_2)$ with the factor (\citealt{KA87.1}, see also \citealt{ZA96.1,MA00.1}) $1 + 2\beta(\overline{z})/3 + \beta^2(\overline{z})/5$, with $\beta(z) \equiv f(z)/b_\mathrm{eff}(z)$ and
\begin{equation}
f(z) \equiv -\frac{d \ln D_+(z)}{d \ln (1+z)}.
\end{equation}
In models with a cosmological constant, the function $f(z)$ can be well approximated by \citep{LA91.1}
\begin{equation}
f(z) \simeq \Omega_\mathrm{m}^{0.6}(z) + \frac{\Omega_\mathrm{x}(z)}{70} \left[ 1 + \frac{\Omega_\mathrm{m}(z)}{2} \right]\,.
\end{equation}
In more general models with a dynamical evolution in the dark-energy component, $f(z)$ must be evaluated numerically.

Taking all past light cone effects into account, the observed spatial correlation function is given by
\begin{equation}
\xi_\mathrm{obs}(r) = \frac{1}{A^2} \int_{z_\mathrm{inf}}^{z_\mathrm{sup}}\int_{z_\mathrm{inf}}^{z_\mathrm{sup}} \frac{\mathcal{N}(z_1)}{r(z_1)}  \frac{\mathcal{N}(z_2)}{r(z_2)} \xi_\mathrm{obj}(r,z_1,z_2) dz_1dz_2\,,
\end{equation}
with normalisation
\begin{equation}
A \equiv \int_{z_\mathrm{inf}}^{z_\mathrm{sup}} \frac{\mathcal{N}(z)}{r(z)} dz\,.
\end{equation}
In the previous two equations, $z_\mathrm{inf}$ and $z_\mathrm{sup}$ are the minimum and maximum redshifts, respectively spanned by the cluster catalog at hand. In realistic situations, $z_\mathrm{inf} \simeq 0$, while $z_\mathrm{sup}$ depends on the sensitivity of the instrument used.

Accordingly, the observed angular correlation function is
\begin{equation}\label{eqn:angular}
\omega_\mathrm{obs}(\theta) = \frac{1}{B^2} \int_{z_\mathrm{inf}}^{z_\mathrm{sup}}\int_{z_\mathrm{inf}}^{z_\mathrm{sup}} \mathcal{N}(z_1)  \mathcal{N}(z_2) \xi_\mathrm{obj}(\overline{r},z_1,z_2) dz_1dz_2,
\end{equation}
where

\begin{equation}
\overline{r} = \overline{r}(z_1,z_2,\theta) \equiv \sqrt{r^2(z_1) + r^2(z_2) - 2r(z_1)r(z_2)\cos(\theta)}
\end{equation}
and the normalisation in this case is the total number of objects included in the catalog at hand,

\begin{equation}
B \equiv \int_{z_\mathrm{inf}}^{z_\mathrm{sup}} \mathcal{N}(z) dz. 
\end{equation}

In the small angle approximation (see for instance \citealt{PE80.1}), the relation in Eq.~(\ref{eqn:angular}) simplifies to

\begin{equation}
\omega_\mathrm{obs}(\theta) = \frac{1}{B^2} \int_{z_\mathrm{inf}}^{z_\mathrm{sup}} \frac{\mathcal{N}^2(z)}{dr(z)/dz}
\int_{-\infty}^{+\infty} \xi_\mathrm{obj}(r_*,z,z) du dz,
\end{equation}
with

\begin{equation}
r_* = r_*(u,\theta,z) \equiv \sqrt{u^2 + r^2(z)\theta^2}.
\end{equation}

With the help of this formalism, it is possible to produce realistic theoretical expectations for the correlation properties of galaxy clusters in the past light cone, as a function of catalog properties and the cosmological model.

\section{Survey properties}\label{sct:survey}

We consider five ongoing and planned cluster surveys here, two of which are X-ray based, while the remaining three are SZ based.

\subsection{X-ray catalogs}

Two forthcoming X-ray surveys are addressed in this work. The first has the properties of the \emph{eRosita} wide survey, described in the mission definition document \footnote{available at \\\texttt{http://www.mpe.mpg.de/projects.html\#erosita}}. It is designed to have a sky coverage of $f_\mathrm{sky} \simeq 0.485$ and a limiting flux of $F_\mathrm{lim} = 3.3 \times 10^{-14}$ erg s$^{-1}$ cm$^{-2}$ in the energy band $\left[  0.5,2.0 \right]$ keV. We note that this planned survey fulfills almost exactly the requirements specified in the dark-energy task force paper of \cite{HA05.1}, where a survey of X-ray clusters optimal to constraining the evolution in the dark-energy equation-of-state is described. This proposed survey covers $\Omega \simeq 2 \times 10^4$ square degrees ($f_\mathrm{sky} \simeq 0.485$) like the \emph{eRosita} wide survey, and has a limiting flux in the $\left[  0.5,2.0 \right]$ keV energy band that is only slightly lower of, $F_\mathrm{lim} = 2.3 \times 10^{-14}$ erg s$^{-1}$ cm$^{-2}$. We adopt however the exact parameters of the \emph{eRosita} survey.

The second is the XMM cluster survey (\citealt{SA08.1}, XCS henceforth), a serendipitous search for X-ray clusters in the existing exposures of the XMM satellite archive. The sky coverage estimated on the basis of the already surveyed area, the pointings still to be analysed, and the expected mission lifetime is $\Omega \simeq 500$ square degrees ($f_\mathrm{sky} \simeq 0.012$). As for the depth of the survey, since the XMM archive occupies a range of different exposure times, and thus of limiting fluxes, a single limiting flux is inappropriate for describing the XCS catalog. If fluxes in the $\left[  0.1,2.4 \right]$ keV energy band are considered, a single limiting flux of $F_\mathrm{lim} = 3.5 \times 10^{-13}$ erg s$^{-1}$ cm$^{-2}$ might be used, although the redshift-dependent cut reported in Fig.~9 of \cite{SA08.1} is more appropriate, being fairly fit by

\begin{equation}
\frac{F_\mathrm{lim}(z)}{10^{-13} \mbox{ erg s}^{-1}\mbox{ cm}^{-2}} = 2.8\: z^{-1/3}.
\end{equation}

We note that the XMM cluster survey has a brighter limiting flux at all cosmological redshifts and a much smaller covered area than the \emph{eRosita} survey, hence large differences are expected in the number of clusters,  and the clustering signal-to-noise ratio detected in the former survey will be much lower than in the latter.

\subsection{SZ catalogs}

We consider three planned blind sub-mm surveys. The first is based on South Pole Telescope (SPT) observations, and was considered by \cite{MA03.1} in an attempt to predict possible constraints on cosmological parameters and dark energy. The proposed survey area amounts to $\Omega \simeq 4 \times 10^3$ square degrees ($f_\mathrm{sky} \simeq 0.097$), for a limiting SZ flux density (see definition in the Sect. \ref{sct:scaling} below) of $S_{\nu_0,\mathrm{lim}} \simeq 5$ mJy at a frequency $\nu_0 \equiv 150$ GHz. We note that a blind survey of SZ clusters with SPT has indeed already started with the first successful detections \citep{ST08.1}, hence a comparison of our theoretical predictions with observations may be imminent.

The second SZ cluster survey is based on the portion of sky that the Atacama Cosmology Telescope (ACT) will observe. According to \cite{SE07.1}, this will consist of two stripes covering 4 degrees in declination and 360 degrees in right ascension, for a total of $\Omega \simeq 3.6 \times 10^3$ square degrees ($f_\mathrm{sky} \simeq 0.087$). The depth of the survey is defined in terms of the (frequency-independent) integrated Compton-$y$ parameter over the solid angle covered by the virial sphere of each individual cluster, $Y_{200}$. According to the simulations performed in \cite{SE07.1}, for a limiting integrated Compton parameter of $Y_{200,\mathrm{lim}} \simeq 10^{-3}$ arcmin$^2$, a galaxy cluster-sample $\sim 90\%$ complete is produced, even with the inclusion of noise from radio and infrared point sources.

The last survey that we consider will be carried out by the \emph{Planck} satellite. This is the largest sub-millimeter survey of the sky that is currently being developed. Even though the beam of the satellite will be quite large, it is predicted to detect several thousands of clusters by their thermal SZ effect. As shown in \cite{SC07.1}, the sky coverage for the detection of galaxy clusters will be highly non-uniform, especially at low detection significance. On the other hand, we believe that a uniform sky coverage of $\Omega \simeq 3 \times 10^{4}$ square degrees ($f_\mathrm{sky} \simeq 0.727$) is realistic and sufficient for our purposes, and hence we adopt it here. As for the limiting SZ observable, the noise due to Planck's scanning path is highly structured on cluster-scales and shorter length. This means that assuming a simple flux-detection threshold is insufficient for our study. In \cite{SC07.1}, the minimum mass detected as a function of redshift is presented for a limiting integrated $y$-parameter of $Y_{200,\mathrm{lim}} = 10^{-3}$ arcmin$^2$, based on a numerical simulation in a $\Lambda$CDM world model. This minimum mass is virtually independent of the filtering scheme adopted, and is described reasonably well by

\begin{equation}
\log \left(\frac{M_\mathrm{lim}(z)}{10^{15} M_\odot h^{-1}}\right) = -1.200 + 1.469\arctan\left[ \left(z-0.10\right)^{0.440} \right]
\end{equation}
for $z \ge 0.11$, and by 

\begin{equation}
\log \left(\frac{M_\mathrm{lim}(z)}{10^{15} M_\odot h^{-1}}\right) = -1.924 + 8.333 z 
\end{equation}
if $z \le 0.11$.
This kind of fit may appear cumbersome, but we were unable to find a simpler functional form that adequately reproduces the results of \cite{SC07.1}, because of the steep increase in the limiting mass at $z \gtrsim 0.1$. Note that the two branches of the fit join smoothly at $z = 0.11$.

The simulation by \cite{SC07.1} was performed for a single cosmological model, namely a WMAP-1 cosmology \citep{SP03.1} with $\Omega_{\mathrm{m},0} = 0.3$, $\Omega_{\Lambda,0} = 0.7$, $h = 0.7$ and $\sigma_8 = 0.9$. To correct for the different cosmologies used here, we proceed as follows. First, we convert the minimum mass from \cite{SC07.1} into a minimum integrated Compton $y$-parameter, according to the scaling relation described in Sect.~\ref{sct:scaling} below and by considering the appropriate WMAP-1 cosmology. We then convert this minimum integrated $y$-parameter back into a minimum mass using the same scaling relation but using the various cosmologies adopted here. This procedure explains why clusters of the same mass produce different signals in different cosmologies, because of their different formation histories and the differences in geometry of the universe between models. On the other hand, altering the cluster abundance may also change the amount of undetected objects, and thus the background noise. A proper treatment of this issue would require a far more detailed analysis and probably fully numerical simulations, which we decide was unnecessary for our purposes. 

\section{Scaling relations}\label{sct:scaling}

To relate survey properties to the extent in mass and redshift space of the resulting cluster catalog, it is necessary to link the mass and redshift of an individual cluster to the relevant observable, namely X-ray flux for X-ray surveys and SZ flux density or integrated Compton-$y$ parameter for sub-mm surveys. We do this by means of realistic scaling laws. We note that not all features of these scaling relations are well established, especially concerning their redshift evolution. We described in the following what we propose to be the most suitable way to proceed given our aims.

\subsection{X-ray scaling relations}

First of all, we used the conversion between the X-ray temperature and the virial mass adopted in \cite{FE07.2} (see also \citealt{BA03.1}), i.e.,~a virial relation with normalisation based on the simulations of \cite{MA01.1}

\begin{equation}\label{eqn:mt}
kT(M_{200},z) = 4.88 \mbox{ keV} \left[ \frac{M_{200}}{10^{15} M_\odot} h(z) \right]^{2/3},
\end{equation}
where the mass is measured in units of $M_\odot$. Additionally, a luminosity-temperature relation given by

\begin{equation}\label{eqn:lt}
L(T) = 2.5 \times 10^{43} \mbox{ erg s}^{-1} h^{-2} \left( \frac{kT}{1.66 \mbox{ keV}} \right)^{2.331}
\end{equation}
was used. This is based on observations by \cite{AL98.1}, and is assumed not to evolve with redshift according to the analyses of \cite{MU97.1}, \cite{RE99.1} and \cite{HA02.1}. Combining these two relations, we obtain the mass-luminosity scaling law

\begin{equation}\label{eqn:lm}
L(M_{200},z) = 3.087 \times 10^{44} \mbox{ erg s}^{-1} h^{-2} \left[ \frac{M_{200}}{10^{15} M_\odot} h(z)\right]^{1.554}.
\end{equation}

Choosing a reasonable value for the Hubble constant, $h = 0.7$, Eq.~(\ref{eqn:lm}) equals

\begin{equation}
L(M_{200},z) = 1.097 \times 10^{45} \mbox{ erg s}^{-1} \left[ \frac{M_{200}}{10^{15} M_\odot h^{-1}} E(z) \right]^{1.554},
\end{equation}
where the mass is now expressed in $M_\odot h^{-1}$. \cite{BA03.1} demonstrated that this mass-luminosity relation is a good fit to the X-ray cluster observations compiled by \cite{RE02.1}.

A possible steepening of the luminosity-temperature relation for low-mass clusters and groups of galaxies has also been advocated \citep{HE00.2,HE00.1,XU00.1}, which would require replacing Eq.~(\ref{eqn:lt}) with a broken power-law. However, it has been shown \citep{OS04.1,KH07.1} that the scaling relation for groups is consistent with that for clusters, although the scatter for groups is considerably larger. This could bias the estimate of the relation's slope; However since this has not been established definitively, we prefer to adhere to Eq.~(\ref{eqn:lt}) in the following. A steepening of the luminosity-temperature relation for groups of galaxies would have the consequence of including fewer low-mass objects in the various catalogs. These would thus contain a higher fraction of high-mass clusters whose clustering properties would enhance the differences between cosmological models. Our results may thus slightly underestimate the distinguishing power of the correlations.

To convert the bolometric X-ray luminosity provided by the scaling relations to the luminosity in a given band required to characterize the cluster catalogs of \emph{eRosita} and XCS, we adopt a Raymond-Smith \citep{RA77.1} plasma model implemented via the \texttt{xspec} software package \citep{AR96.1}, with metal abundance $Z = 0.3 Z_\odot$ \citep{FU98.1,SC99.1}. Once the mass-luminosity relation is obtained, the mass (and redshift)-flux relation trivially follows.

\subsection{SZ scaling relations}

The SZ effect is a scattering process appearing in the CMB spectrum as absorption at frequencies below $\sim 218$ GHz and as emission above. Nonetheless, a flux density can be formally associated with the temperature distortion imprinted by the thermal SZ effect in the following way.

We consider the Compton-$y$ parameter observed in a given direction $\theta$ of the sky and integrate it over an arbitrary solid angle $\Omega$ to obtain

\begin{equation}
Y = \int_\Omega y(\theta) d^2\theta.
\end{equation}
The temperature distortion over the patch of the sky covered by $\Omega$ is proportional to $Y$ times the typical frequency pattern of the thermal SZ effect, hence the monochromatic SZ flux per unit frequency received from the solid angle $\Omega$ can be defined as $S_\nu \equiv j(\nu) Y$, with

\begin{equation}
j(\nu) = 2 \frac{(kT_\gamma)^3}{(hc)^2} \left| f(\nu) \right|.
\end{equation}
Here, $T_\gamma$ is the CMB temperature and $f(\nu)$ is the typical spectral signature of the thermal SZ effect,

\begin{equation}
f(\nu) = \frac{x^4 e^x}{(e^x-1)^2} \left[ x\frac{e^x+1}{e^x-1} -4 \right],
\end{equation}
where $x \equiv h\nu/k T_\gamma$, and relativistic corrections are ignored (see however \citealt{IT04.1} and references therein). 

In \cite{SE07.1}, a prescription for linking the mass of a cluster to the Compton parameter integrated over the solid angle subtended by the virial sphere is proposed based on numerical simulations, according to

\begin{equation}\label{eqn:yint}
Y_{200}(M_{200},z) = \frac{2.504 \times 10^{-4}}{\left( D_\mathrm{A}(z)/1 \mbox{ Mpc} \right)^2} \left(\frac{M_{200}}{10^{15} M_\odot}\right)^{1.876}E(z)^{2/3}.
\end{equation}
If we wish to convert the relation in Eq.~(\ref{eqn:yint}) into a scaling law for the SZ equivalent flux density, the slope and redshift dependence will obviously remain unchanged. The normalization, however, must be converted into an SZ flux. We shall assume that the SZ monochromatic flux relates to $Y_{200}$, in the sense that the amount of SZ signal detected outside the virial radius is negligible. Keeping in mind that

\begin{equation}
2 \frac{(kT_\gamma)^3}{(hc)^2} = 2.701 \times 10^{-18} \mbox{ J m}^{-2} = 2.701 \times 10^{11} \mbox{ mJy},
\end{equation}
the scaling law of Eq.~(\ref{eqn:yint}) turns into

\begin{equation}
S_\nu(M_{200},z) = \frac{6.763 \times 10^{7} \mbox{ mJy}}{\left(D_\mathrm{A}(z)/1 \mbox{ Mpc}\right)^2} \left(\frac{M_{200}}{10^{15} M_\odot}\right)^{1.876} \left| f(\nu) \right| E(z)^{2/3}.
\end{equation}
As stated in Sect.~\ref{sct:survey}, for the SPT catalog construction we assumed $\nu = \nu_0 \equiv 150$ GHz, for which $f(\nu_0) = -3.833$. The negative value indicates absorption, as is to be expected since $\nu_0 < 218$ GHz. It follows

\begin{equation}
S_{\nu_0}(M_{200},z) = \frac{2.592 \times 10^{8} \mbox{ mJy}}{\left(D_\mathrm{A}(z)/1 \mbox{ Mpc}\right)^2} \left(\frac{M_{200}}{10^{15} M_\odot}\right)^{1.876} E(z)^{2/3}.
\end{equation}
We use this scaling relation for the SPT cluster catalog.

\section{Catalog properties}\label{sct:cat}

\begin{figure}[t]
  \includegraphics[width=\hsize]{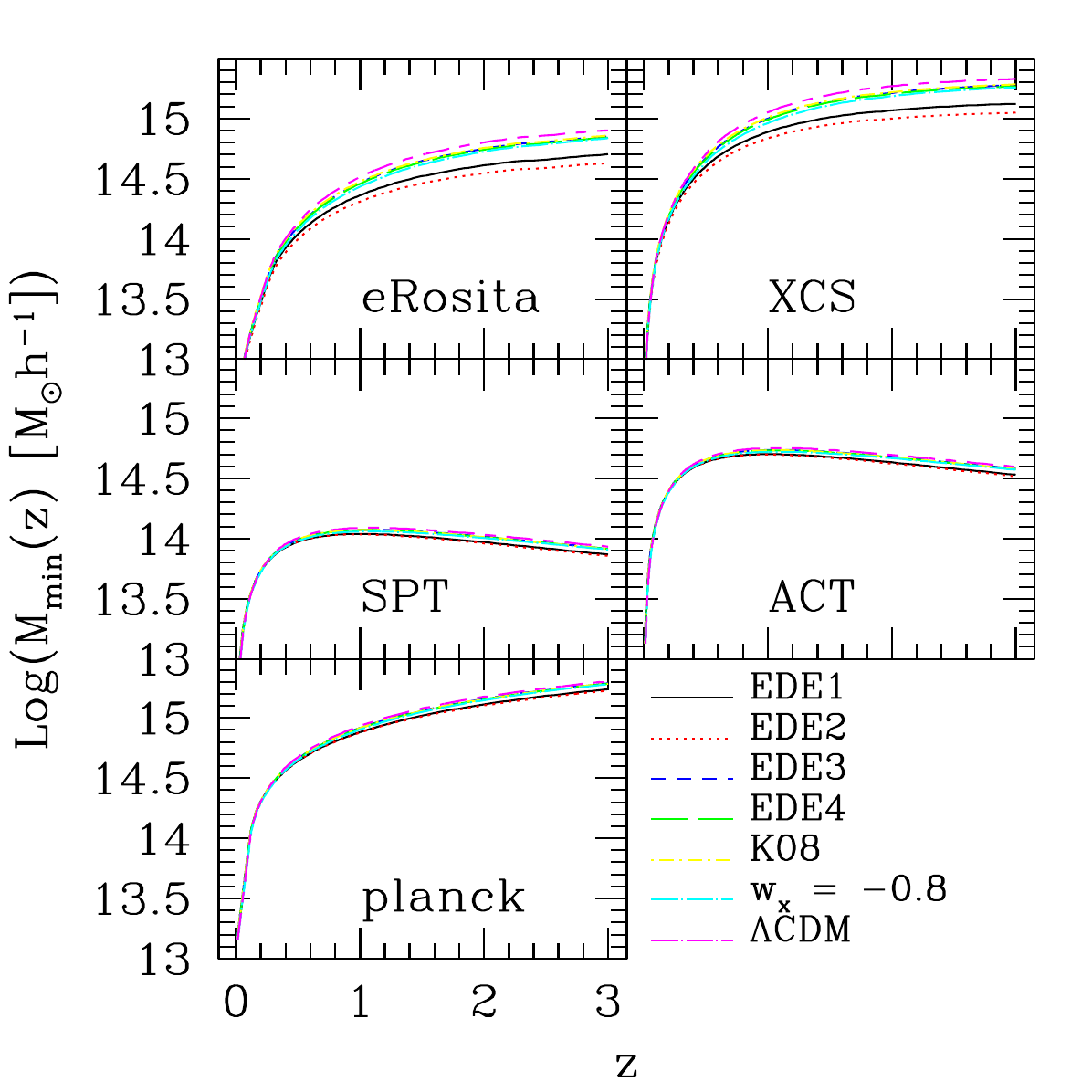}\hfill
\caption{The minimum mass for clusters included in the catalogs constructed with the five different surveys investigated in this work, as labelled in the plot. Results for the seven different dark-energy models described in the text are shown, using the same color and line types as in Fig.~\ref{fig:wz}.}
\label{fig:min}
\end{figure}

Figure~\ref{fig:min} shows the minimum mass as a function of redshift for a cluster to enter each catalog, for all the dark-energy models employed in this work. To compute that, we simply converted the limiting flux or integrated Compton-$y$ parameter for the various surveys into a mass by means of the scaling relations described in Sect.~\ref{sct:scaling}. We notice that, as expected, the minimum mass for X-ray selected catalogs is a monotonically increasing function of redshift, while this is not the case for SZ catalogs, whose minimum mass slightly decreases at high redshift. This because the intrinsic redshift-independence of the SZ decrement causes the SZ flux density or integrated Compton $y$-parameter to scale as the inverse of the angular-diameter distance squared, while the X-ray flux scales as the inverse of the luminosity distance squared, and the former tends to flatten at high redshift. An exception to this behavior occurs for \emph{Planck}, whose limiting mass behaves in a more similar way to the X-ray catalogs because \emph{Planck} has a very large beam that significantly smoothes the signal, especially when the angular size of the source is small.

The differences between different cosmologies arise because the scaling relations used in this work include distances and the expansion history of the universe, and hence depend on cosmology. The justification for this is that in models with dynamical dark-energy, and in early dark-energy models in particular, structure formation begins at earlier times than in more standard models with $w_\mathrm{x} = $ constant. As a consequence, clusters at a given redshift have more concentrated host dark-matter halos and more compact gas distributions, which enhances the SZ effect and X-ray emission. This agrees with the differences in the minimum mass for a given cosmology being more pronounced for X-ray catalogs, because X-ray emission is proportional to the square of the gas density, while the SZ effect scales only linearly with the density. 

Among the X-ray cluster catalogs, XCS has a systematically higher minimum mass compared to \emph{eRosita}, because of the higher minimum observed flux. Similarly, ACT has a higher minimum integrated Compton $y-$parameter than SPT, and hence the minimum mass included in the catalog is systematically higher. The minimum mass for the \emph{Planck} catalog is always much larger than that of SPT and ACT, thus we expect the number of clusters per unit area entering in its catalog to be relatively small. However, this is in some way compensated by the large area of the \emph{Planck} survey.

\begin{figure}[t]
  \includegraphics[width=\hsize]{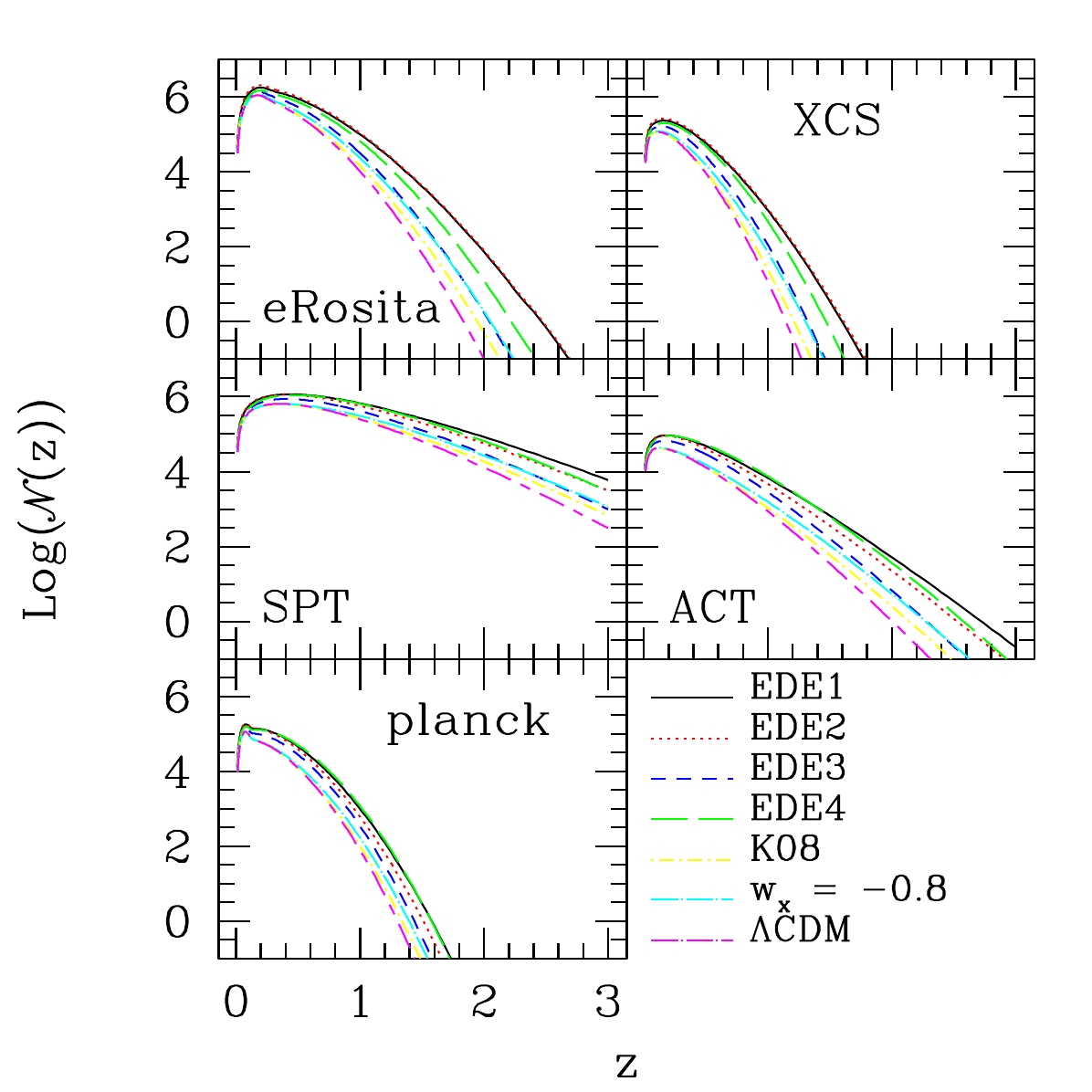}\hfill
\caption{The redshift distribution (all-sky equivalent) of clusters entering each of the five catalogs used in this work. Differences between different cosmologies are shown with the same color and line types as in previous figures, as labelled in the plot.}
\label{fig:dist}
\end{figure}

The redshift distribution of objects entering the various cluster catalogs for the seven different cosmologies used in this work is shown in Fig.~\ref{fig:dist}. As expected, large differences occur between different survey catalogs and different dark-energy models. The models EDE1 and EDE2 show very similar results, and also the largest number of objects included in a given catalog. This is due to the enhanced cluster abundance in these models \citep{FE07.1}. The model EDE3 shows results that are very similar to the model with constant $w_\mathrm{x} = -0.8$, except for a moderate difference at low redshifts, close to the peak of the distribution. This is agrees with \cite{WA08.1}, where it was shown that the cluster number counts predicted to be obtained with \emph{Planck} in model EDE3 is almost identical to those for a standard $\Lambda$CDM model. The $\Lambda$CDM model always produces the smallest number of objects, for all catalogs and at all redshifts, with the K$08$ model being intermediate between the $\Lambda$CDM and the $w_\mathrm{x} = -0.8$ models.

As expected, the SZ surveys (except for \emph{Planck}, which has a minimum mass behavior more similar to the X-ray catalogs) have a far wider redshift distribution than the X-ray catalogs. This is because the minimum mass slightly decreases at high redshift, as opposed to a monotonic increase (see Fig.~\ref{fig:min}), and the sample may contain a high number of low-mass objects. For SPT we  caution that the redshift distribution remains significantly above zero at the limiting redshift of our analysis, $z = 3$. However, as we verified, the number of objects included in this catalog with $z \ge 3$ is just $\sim 0.4\%$ of the total, and hence negligibly contributes in the redshift integrals needed for computing the observed spatial and angular correlation functions. Those objects might be significant when binning the catalog in redshift, but again, the number of clusters with $z \ge 3$ is at most a few per cent of the number with $z \ge 1.5$, hence we assume that those objects can be safely neglected. The scaling relation between the SZ flux density and the mass of the host dark-matter halo is indeed highly uncertain for low masses and high redshifts, and hence we prefer to cut our sample at $z = 3$.

\section{Results}\label{sct:res}

\subsection{Full catalogs}

\begin{figure}[t]
  \includegraphics[width=\hsize]{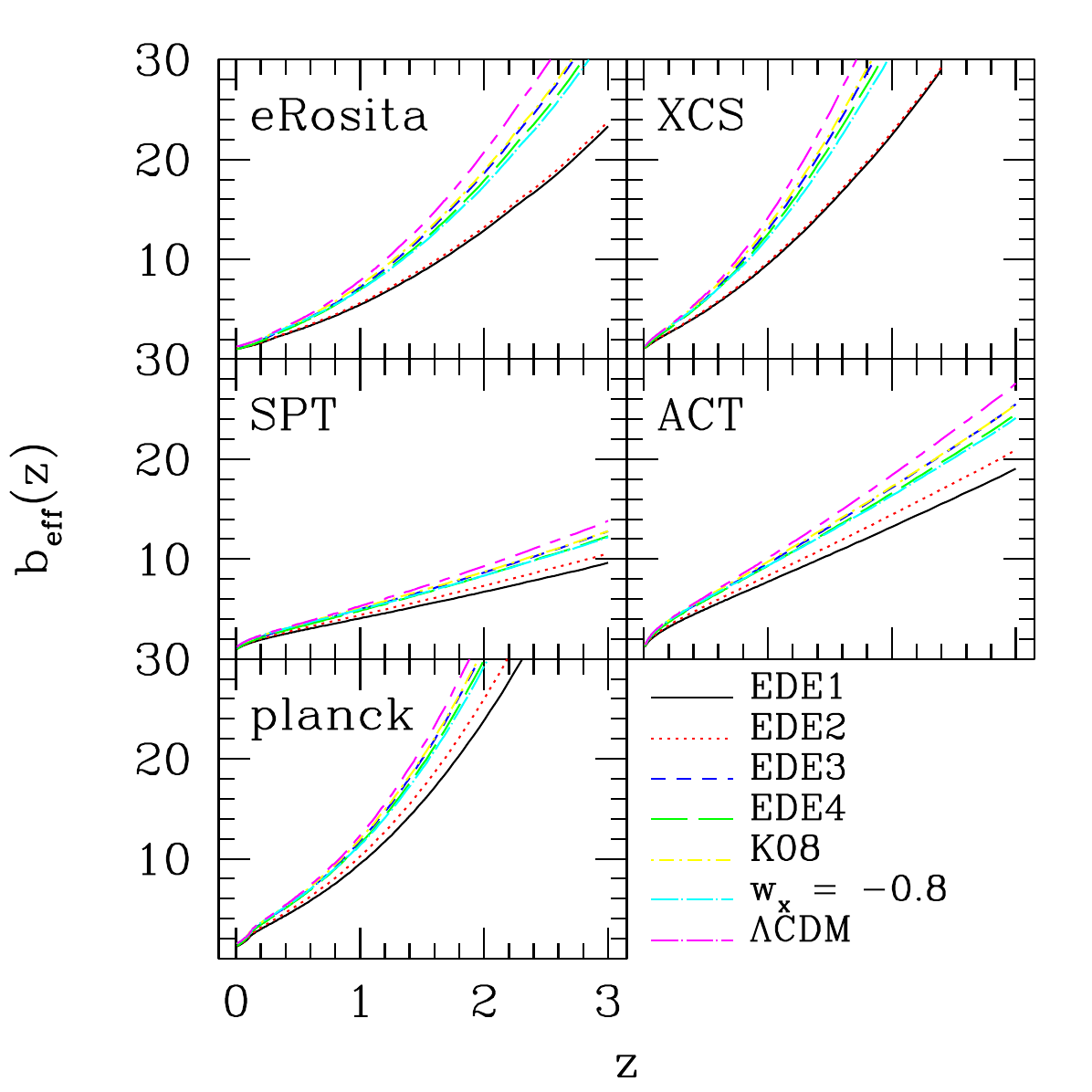}\hfill
\caption{Effective bias for the five cluster catalogs used in this work. Different colors and line styles refer to different cosmological models, as labelled in the plot.}
\label{fig:bias}
\end{figure}

In Fig.~\ref{fig:bias}, we show the effective bias computed by using all the clusters in the different samples investigated in this work. Results for the seven different cosmological models described in Sect.~\ref{sct:cosmo} are also shown. The effective bias for models EDE1 and EDE2 is always significantly smaller than for the other models, and this holds true for all cluster catalogs considered here. This is obviously because forming massive objects is easier in those models, and a much higher abundance of objects is present in the various catalogs than for the other cosmological models. As a consequence, large galaxy clusters are less exceptional objects, and are less biased with respect to the underlying dark-matter density field. Models EDE3 and EDE4 (especially the former) are more similar to the standard $\Lambda$CDM case with respect to bias. For EDE3, this is consistent with the previous discussion and also with the findings of \cite{WA08.1}, while for the EDE4 model this is unexpected, since the cluster abundance at a given redshift (see Fig.~\ref{fig:dist}) is lower than, but quite similar to, those for EDE1 and EDE2. However, model EDE4 has an extremely low normalisation for the power spectrum of linear density fluctuations, which is expected to produce a higher `monochromatic' bias. This is likely to play a major role in the computation of the effective bias.

In line with the previous discussion, the $\Lambda$CDM model produces the largest bias, while the K08 and $w_\mathrm{x} = -0.8$ models give slightly smaller and very similar results to EDE3 and EDE4. In particular, the bias for the EDE1 model is up to a factor of $\sim 2$ smaller than that for $\Lambda$CDM at high redshift. It is important to note that, even though this is not clearly visible in Fig.~\ref{fig:bias}, the ratio of the effective bias in models EDE3 and EDE4 to that in model $\Lambda$CDM actually \emph{decreases} quite steeply with redshift for $z < 0.2$, and increases again at higher redshift. Conversely, for other EDE models this ratio continously increases. This is due to the peculiar behavior of the dark-energy equation-of-state parameter in models EDE3 and EDE4, and will have important consequences on the effect of binning the cluster catalog in redshift, as discussed in the next subsection.

\begin{figure*}[t]
\begin{center}
  \includegraphics[width=0.75\hsize]{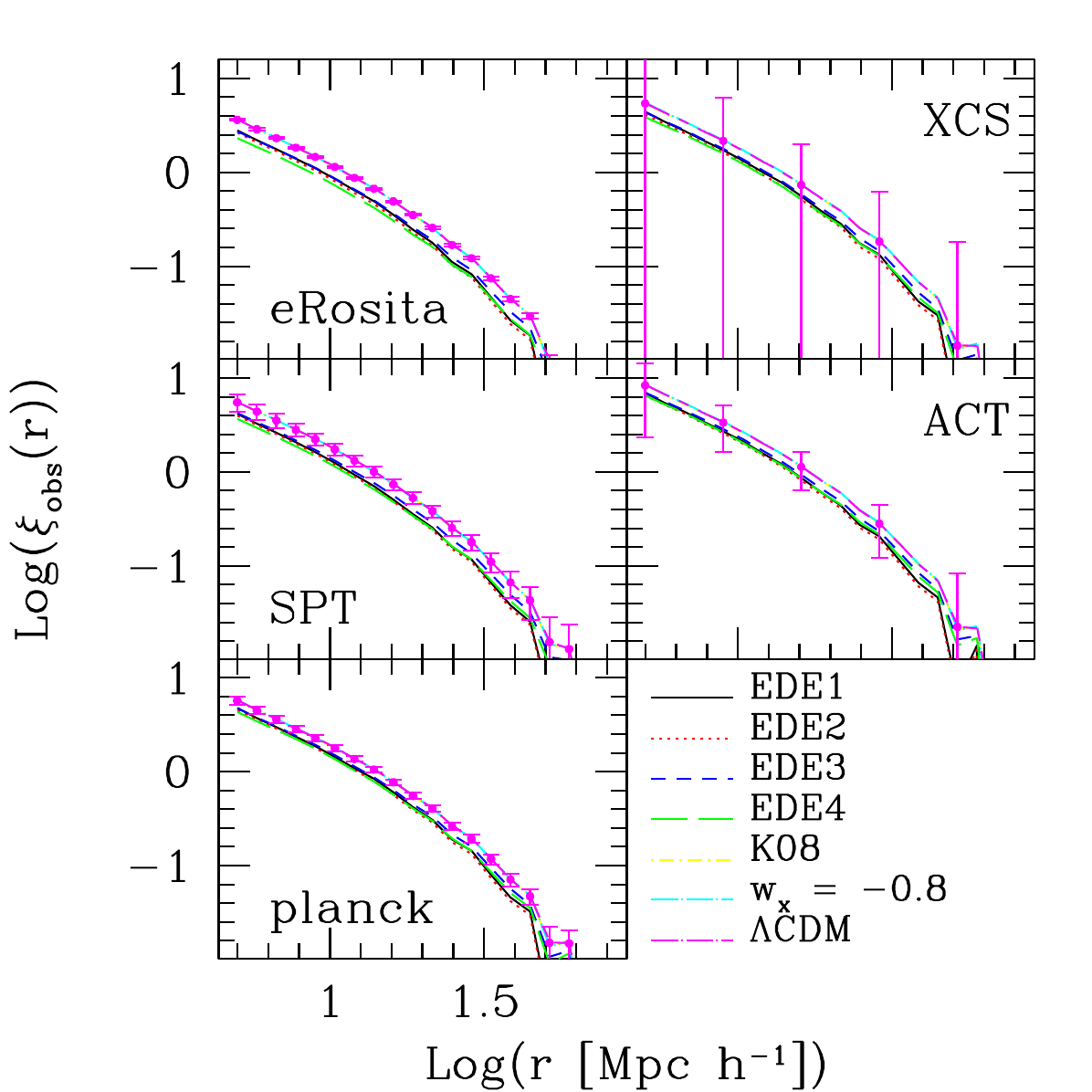}\hfill
\end{center}
\caption{Predicted correlation function for the five cluster catalogs used in this work. Different colors and linestyles refer to different cosmological models, as labelled in the plot. Errorbars are computed with the bootstrap method, and refer only to the $\Lambda$CDM model for clarity.}
\label{fig:correlation}
\end{figure*}

In Fig.~\ref{fig:correlation}, we show the observed correlation function for the five surveys and seven dark-energy models considered here. As expected, the spatial correlation function decreases with increasing radius, and starts oscillating for large separations, $r \gtrsim 40$ Mpc $h^{-1}$. The correlation function in the standard $\Lambda$CDM model is practically indistinguishable from that  in the model with a constant dark-energy equation-of-state parameter $w_\mathrm{x} = -0.8$ and also in the dynamical-dark energy model K08. In both cases, this is due to the similar behavior of the dark-energy equation-of-state parameter (see Fig.~\ref{fig:wz}). In more detail, for the K08 case $w_\mathrm{x}$ differs from the concordance value only at very low redshift, but approaches $-1$ at $z \gtrsim 3$. In particular, the equation of state parameter is identical to the $\Lambda$CDM scenario during the linear stage of structure formation.

On the other hand, models with early dark-energy differ significantly. Because of their lower effective bias, the observed correlation functions are also lower than those for more standard models, by $\sim 50-60\%$ at small radii for \emph{eRosita} and SPT, and slightly less for the other surveys. These observed differences are caused by a combination of different effective bias, linear density-fluctuation correlation functions, and object redshift distributions.

The error bars in Fig.~\ref{fig:correlation} (shown only for the $\Lambda$CDM model for simplicity) were computed with the bootstrap method and imply that the difference between EDE and $\Lambda$CDM-like models would be detectable in the correlation functions observed with \emph{eRosita}, SPT and (to a lesser extent) \emph{Planck}. On the other hand, this difference would be completely lost in the noise for ACT and in particular for XCS. The large error bars visible in the XCS panel are due to the very small sky coverage of this survey, which is not adequately compensated by an increase in depth. This result qualitatively agrees with \cite{MO00.1}, who showed that a deep survey has larger errors in the observed and angular correlation functions than a wide one. The same line of reasoning also applies to ACT, whose limiting integrated Compton-$y$ parameter is too shallow to allow the collection of a significant signal.

We note that in an attempt to reduce the size of the error bars, we increased the radial binning for XCS and ACT, so that the relative error, scaling $\propto 1/\sqrt{dr}$, would decrease. However, even then the size of the error bars remains much larger than the differences in correlation amplitudes between the EDE models and the concordance cosmological scenario. We also note that in principle the same procedure could be applied to the relative errors in the other catalogs, resulting in an even lower amplitude of the relative errors. This is obviously unnecessary in this case, and in the following 
we also perform this operation only for the XCS and ACT catalogs. However, we keep in mind that when the error bars for the other catalogs are only slightly larger than the difference between cosmological models, enlarging the radial binning would probably allow a significant detection of deviations from the concordance $\Lambda$CDM model.

\begin{figure*}[t]
\begin{center}
  \includegraphics[width=0.75\hsize]{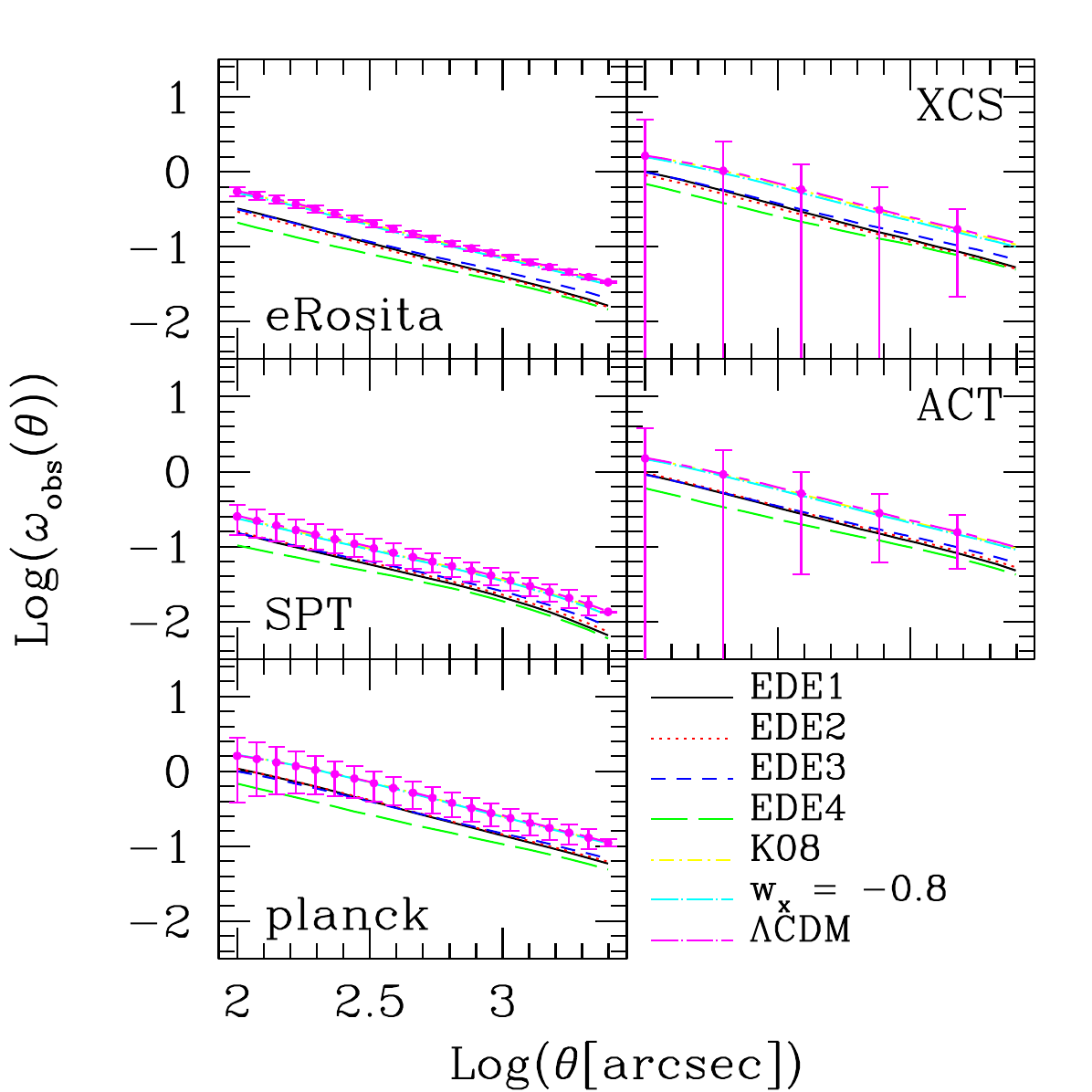}\hfill
\end{center}
\caption{Angular correlation function for the five cluster catalogs used in this work. Different colors and linestyles refer to different cosmological models, as labelled in the plot. Errorbars are computed with the bootstrap method, and refer only to the $\Lambda$CDM model for clarity.}
\label{fig:angular}
\end{figure*}

We note that while \emph{Planck} and XCS have a similar all-sky equivalent redshift distribution of objects, the sizes of the relative errors are very different. This is because of the differences in sky coverage of about three orders of magnitude between the two different surveys, that enter quadratically in the computation of the error bars (see also the comments presented in Sect.~\ref{sct:cat}).

A popular measure of the clustering amplitude from a given cluster catalog is the correlation length, defined as the spatial separation $r_0$ for which the correlation function equals unity, i.e.,~$\xi_\mathrm{obs}(r_0) = 1$. Among the five catalogs analyzed here, the one with the largest correlation length is ACT, having $r_0 \simeq 16$ Mpc $h^{-1}$ for the $\Lambda$CDM cosmology, while the one with the smallest correlation length is \emph{eRosita}, with $r_0 \simeq 10$ Mpc $h^{-1}$ for the same model. As a consequence of the smaller correlation amplitude, the correlation length is also smaller in models with an early dark-energy contribution, by $\sim 20\%$ compared to the concordance cosmology.

If the redshift information about clusters in one catalog is inaccessible or inadequate, i.e.,~if one has only projected information on the plane of the sky, then the only accessible clustering measure is the angular correlation function discussed in Sect.~\ref{sct:clus}. Figure~\ref{fig:angular} shows the angular correlation functions. The differences between different cosmological models and different surveys are enhanced, probably because the angular correlation functions are integrals over the spatial correlation functions along the line-of-sight. This results in a slight separation between the $\Lambda$CDM model and the K08 and $w_\mathrm{x} = -0.8$ models, which is absent from the observed three-dimensional correlation function. Accordingly, in the angular correlation function the ratio of the early-dark energy models to the $\Lambda$CDM-like models can be as high as a factor of $\sim 3$. Finally, differences between individual EDE models are also enhanced, showing that the cosmology producing the lowest angular-correlation amplitude at all separations and for all catalogs is EDE4.

\begin{figure*}[t]
\begin{center}
  \includegraphics[width=0.5\hsize]{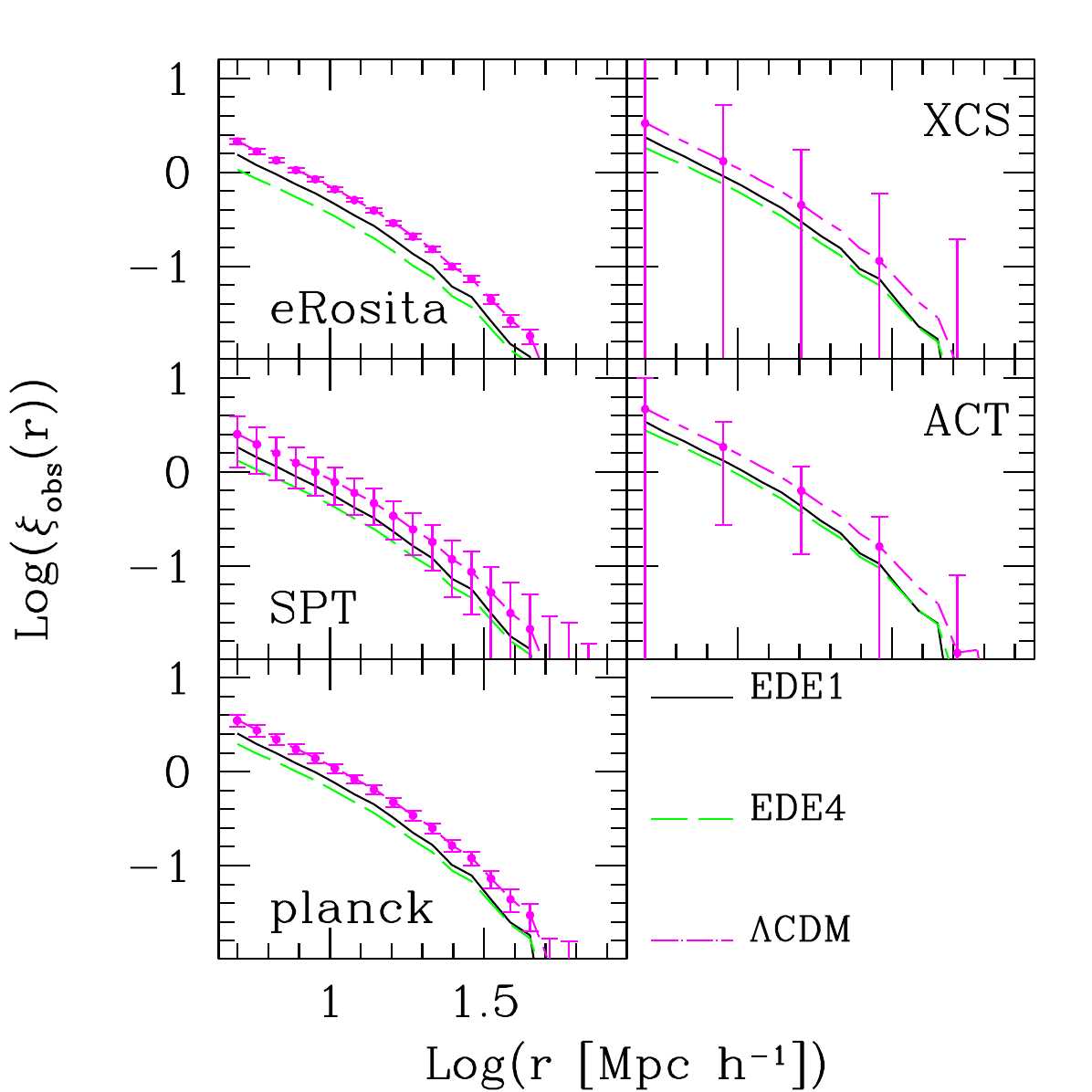}\hfill
  \includegraphics[width=0.5\hsize]{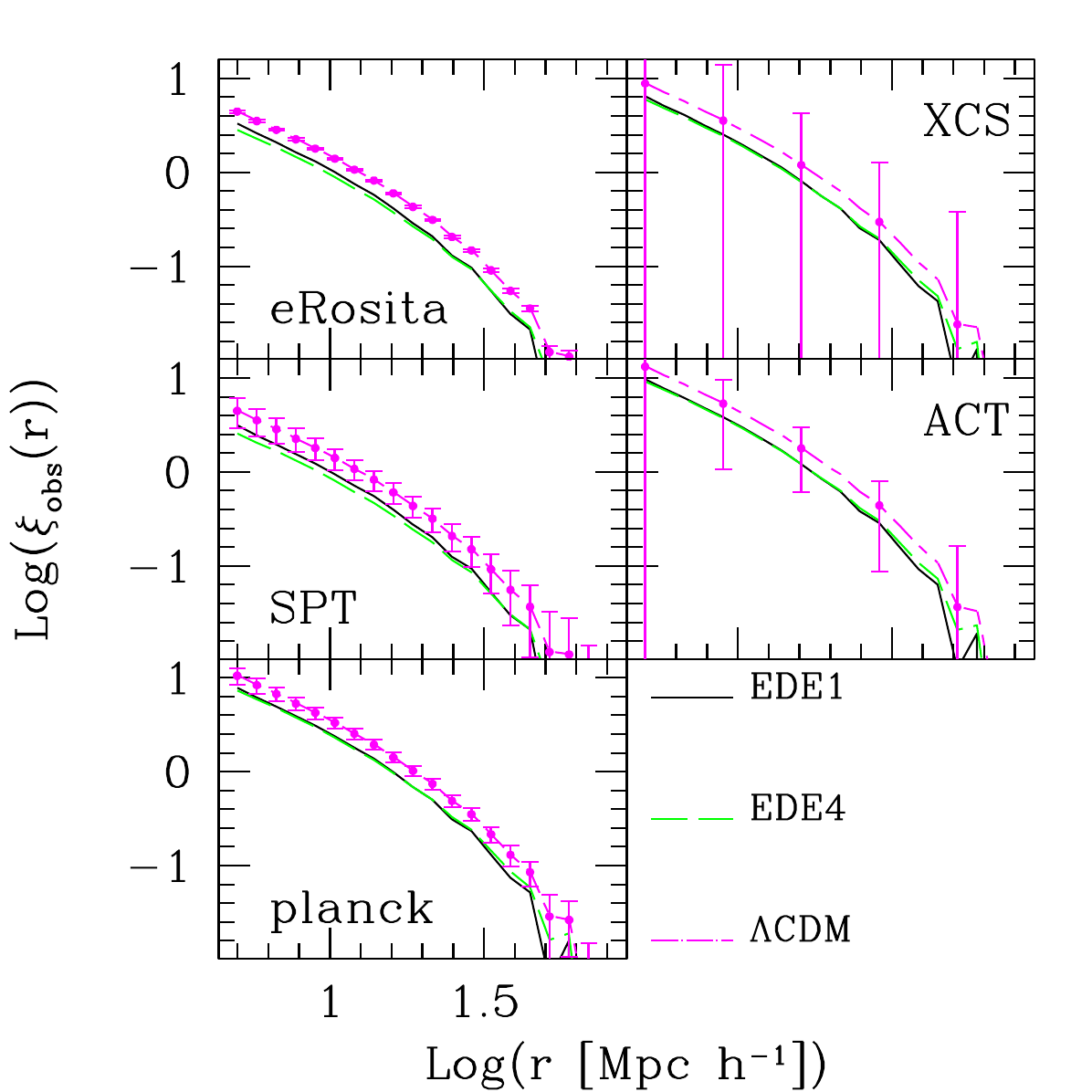}\hfill
\end{center}
\caption{Same as Figure \ref{fig:correlation} but with cluster catalogs restricted to $z \le 0.1$ (left panel) and $z > 0.1$ (right panel). For the SPT catalog, the right panel actually shows the restriction to $0.1 < z \le 0.3$. Only three cosmological models are shown here for simplicity.}
\label{fig:correlation_z_lt}
\end{figure*}

As for the observed spatial correlation function, bootstrap error bars are shown for the $\Lambda$CDM case only for clarity. It is apparent that the difference between $\Lambda$CDM cosmology and EDE models is detected significantly by \emph{eRosita}, while with SPT and \emph{Planck} this is likely only for the most extreme model EDE4, at least with the radial binning we used in the error computation. If this radial binning is increased, the differences between other EDE models and the concordance cosmology become detectable. Finally, as for the spatial correlation function, the error bars for XCS and ACT are too large to allow significant detection of any difference between cosmologies. In both cases, the survey is not deep enough to allow a noteworthy reduction in Poisson noise.

\subsection{Redshift selected catalogs}

In the best-case scenario, knowledge of the redshift of objects entering the catalogs can be assumed. If this is the case, the galaxy-cluster sample can be sub divided into a limited number of redshift bins, such that the number of observed (spatial and/or angular) pairs of objects is broadly the same in each bin. This has the effect of making the relative errors in the correlation functions approximately similar in each bin, and hence allows the coherent study of the redshift evolution in the correlation function. Measuring the redshift evolution in the clustering properties of galaxy clusters on the observational side allows us to determine more accurately the underlying cosmology and dark-energy evolution, since it provides an additional constraint. As a byproduct of this analysis, we can also check whether suitable redshift binning can increase the ratios of the correlation amplitudes in the different cosmological models. 

We note that similar numbers of angular pairs in each redshift bin should correspond to similar absolute number of objects per bin, while this is not necessarily the case for spatial pairs.

We then assume a perfect redshift knowledge of the observed sample. Tests showed that a redshift binning that ensures a suitably high total number of clusters per bin is $z \le 0.1$, $0.1 < z \le 0.3$, and $z > 0.3$ for all catalogs. This choice yields approximately equal signal-to-noise ratios in each bin for the angular correlation function. The same could be obtained for the observed spatial correlation function if each of the bins contained the same number of three-dimensional pairs. This is not true because the highest-redshift bin contains too few spatial pairs compared to the other two. The SPT catalog is the only exception due to its wide redshift distribution (see Fig.~\ref{fig:dist}). Thus, we shall use the same redshift binning for the spatial correlation function only for SPT and two bins, $z \le 0.1$, $z > 0.1$, in all other cases.

\begin{figure}[t]
\begin{center}
  \includegraphics[width=\hsize]{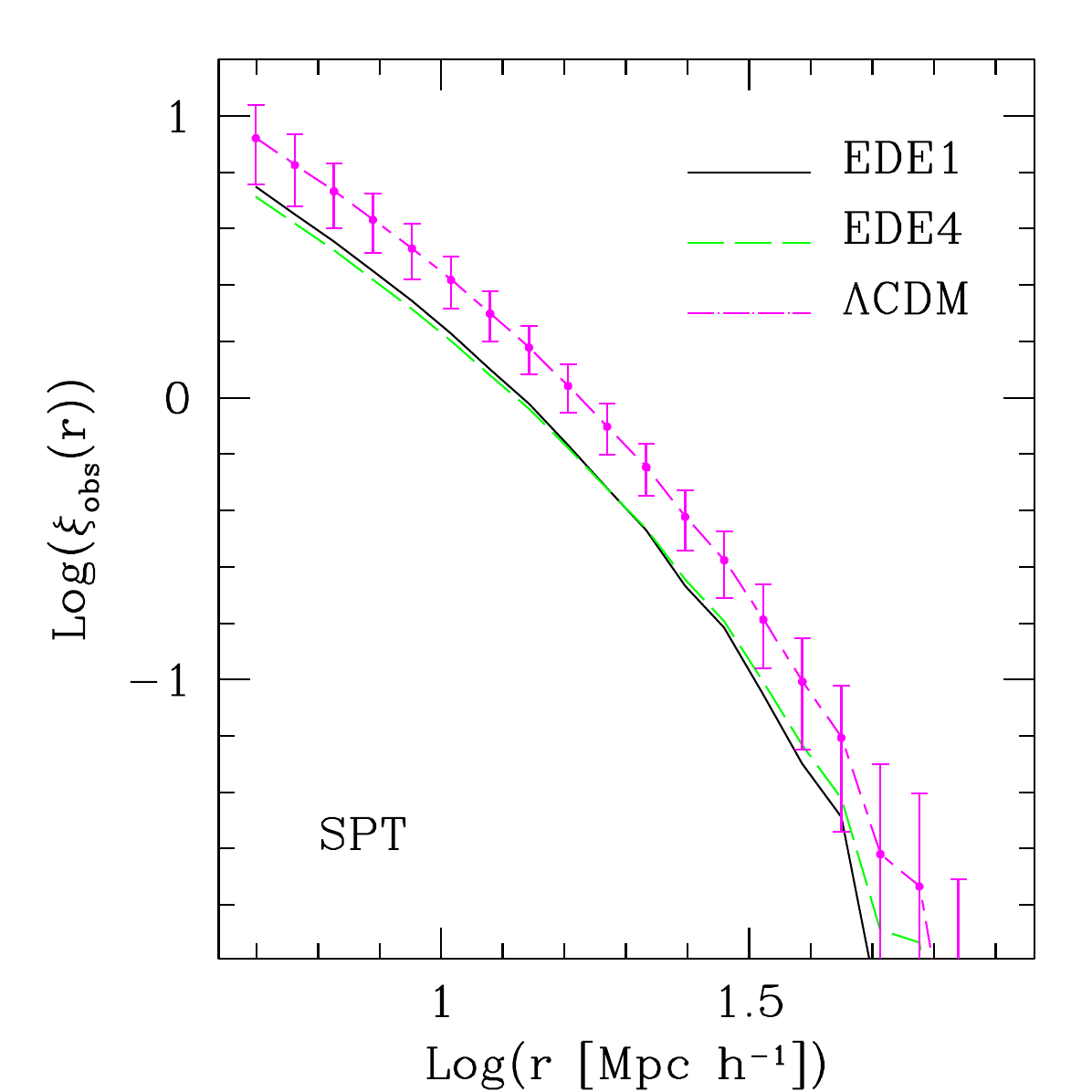}\hfill
\end{center}
\caption{As Fig.~\ref{fig:correlation}, but only for the SPT catalog restricted to $z > 0.3$. Only three cosmologies are shown here for simplicity.}
\label{fig:correlation_z_gt}
\end{figure}

Figure~\ref{fig:correlation_z_lt} shows the spatial correlation function for all cluster catalogs studied here, for $z \le 0.1$ in the left panel and $z > 0.1$ in the right. For the SPT catalog, the right panel shows results for $0.1< z \le 0.3$, while results for $z > 0.3$ are shown in Fig.~\ref{fig:correlation_z_gt}. Here, we consider only three cosmological models, namely the standard $\Lambda$CDM and the two early-dark energy models EDE1 and EDE4. This choice was made because the EDE4 and $\Lambda$CDM models exhibit the largest differences in terms of spatial and angular correlation functions. Additionally, model EDE1 was chosen as being representative of early-quintessence cosmologies with a $w_\mathrm{x}(z)$ evolution that completely differs from EDE4.

The spatial correlation functions computed in all catalogs and for the three different cosmological models increase with increasing redshift, in agreement with the findings of \cite{MO00.1}. The increment is significant for $\emph{eRosita}$ and $\emph{Planck}$, and can become significant for SPT if the radial binning is increased, according to the previous discussion. For example, the correlation length for \emph{Planck} increases from $r_0 \simeq 9$ Mpc $h^{-1}$ to $r_0 \simeq 17$ Mpc $h^{-1}$ in going from $z \le 0.1$ to $z > 0.1$. As for the SPT catalog, for which we added a high-redshift bin, the correlation length increases from $r_0 \simeq 7$ Mpc $h^{-1}$ for $z \le 0.1$ up to $r_0 \simeq 14$ Mpc $h^{-1}$ for $z > 0.3$.

The ratio of the correlation amplitudes in different cosmological models generally increases when clusters are restricted to $z \le 0.1$, and the ratio of the  $\Lambda$CDM and EDE4 models in terms of the spatial correlation function increases significantly. For example, in the \emph{eRosita} sample the ratio of the correlation functions increases from a factor of $\sim 60\%$ to $\sim 2$, and similar increments are seen in the other catalogs. On the other hand, the relative difference between EDE1 and $\Lambda$CDM is practically unchanged compared to the full samples. For high-redshift samples, the ratio of the correlation amplitudes in $\Lambda$CDM and EDE4 models is still larger than for the complete samples, but smaller than for low-redshift samples. The differences between EDE1 and the concordance $\Lambda$CDM models are again very similar to those between the low-redshift and the full samples.

The different behavior of EDE1 and EDE4 can be attributed to the different trend with redshift of the effective bias discussed above. Due to the behavior of the redshift distributions shown in Fig. \ref{fig:dist}, the subsamples with $z \le 0.1$ and $z > 0.1$ are virtually dominated by objects at $z \sim 0.1$ and $z \sim 0.2$ respectively (because of a combination of a sharp decline in the mass function with redshift and volume effects). The ratio of the correlation functions of the underlying density fluctuations in the EDE and $\Lambda$CDM models are virtually unchanged when going from $z \sim 0.1$ to $z \sim 0.2$, and the same is true for the ratio of the effective bias in $\Lambda$CDM and EDE1 model, although in the latter model it increases steadily, but slowly with redshift. Instead, the ratio of the biases in $\Lambda$CDM and EDE4 decreases steeply between $z \sim 0$ and $z \sim 0.2$; The difference between the EDE4 and $\Lambda$CDM model is therefore expected to be more enhanced in low-redshift than high-redshift catalogs, while the difference with EDE1 remain about the same.

The relative errors are still very small for \emph{eRosita} in all redshift-selected catalogs, thus the signal-to-noise ratio for differences in the correlation amplitudes between $\Lambda$CDM and EDE models is optimal for the $z \le 0.1$ catalog. The same is also true for \emph{Planck}, because the error bars tend to increase in the $z > 0.1$ sample. For SPT, the only redshift bin in which a difference between the $\Lambda$CDM and the EDE1 or EDE4 models can be reliably detected is the high-redshift bin, $z > 0.3$. For this bin, we obtain essentially the same results as for the full sample. The situation remains practically unchanged for XCS and ACT, where the relative errors are still very large and do not allow any significant detection of the differences between concordance and more exotic models.

\begin{figure*}[t]
\begin{center}
  \includegraphics[width=0.5\hsize]{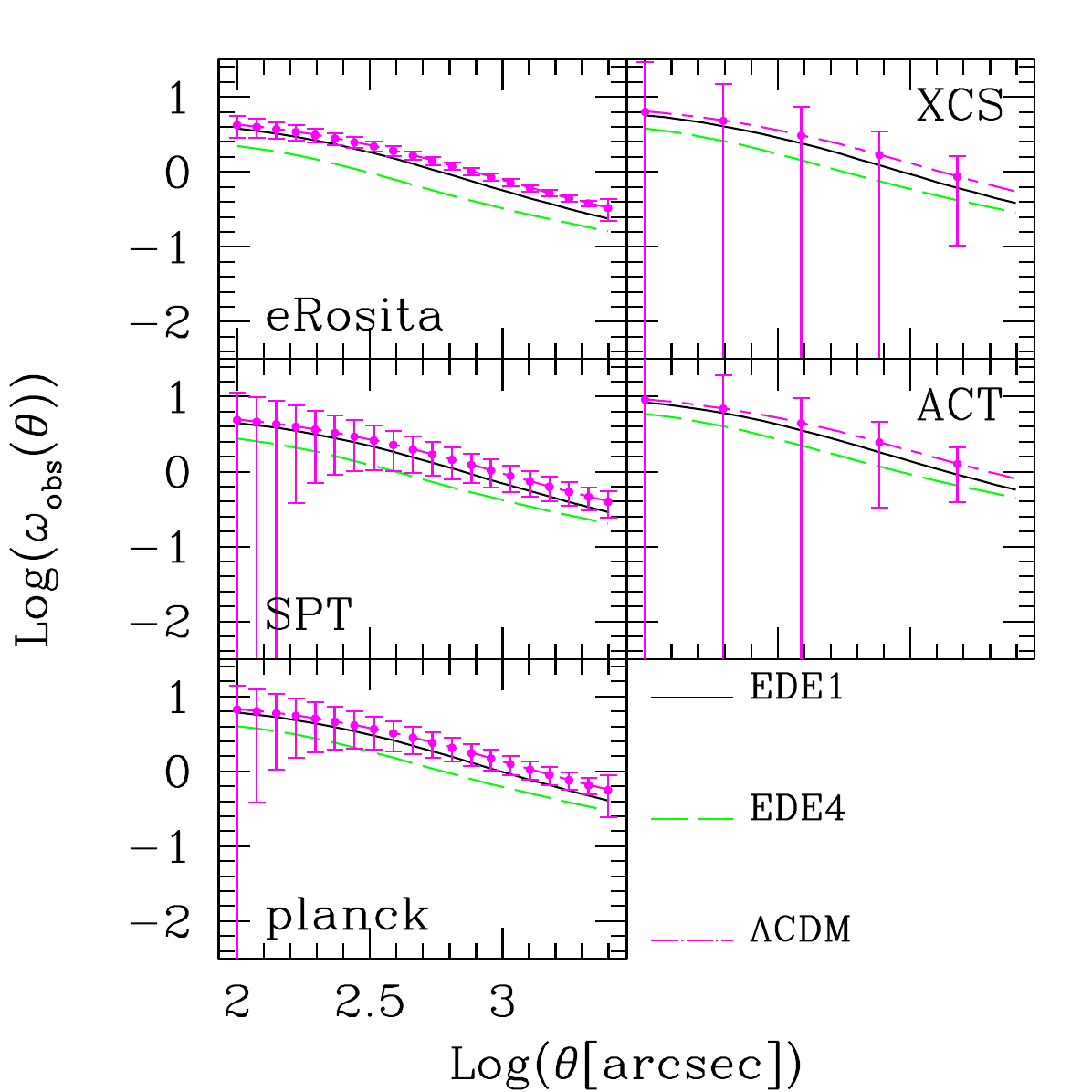}\hfill
  \includegraphics[width=0.5\hsize]{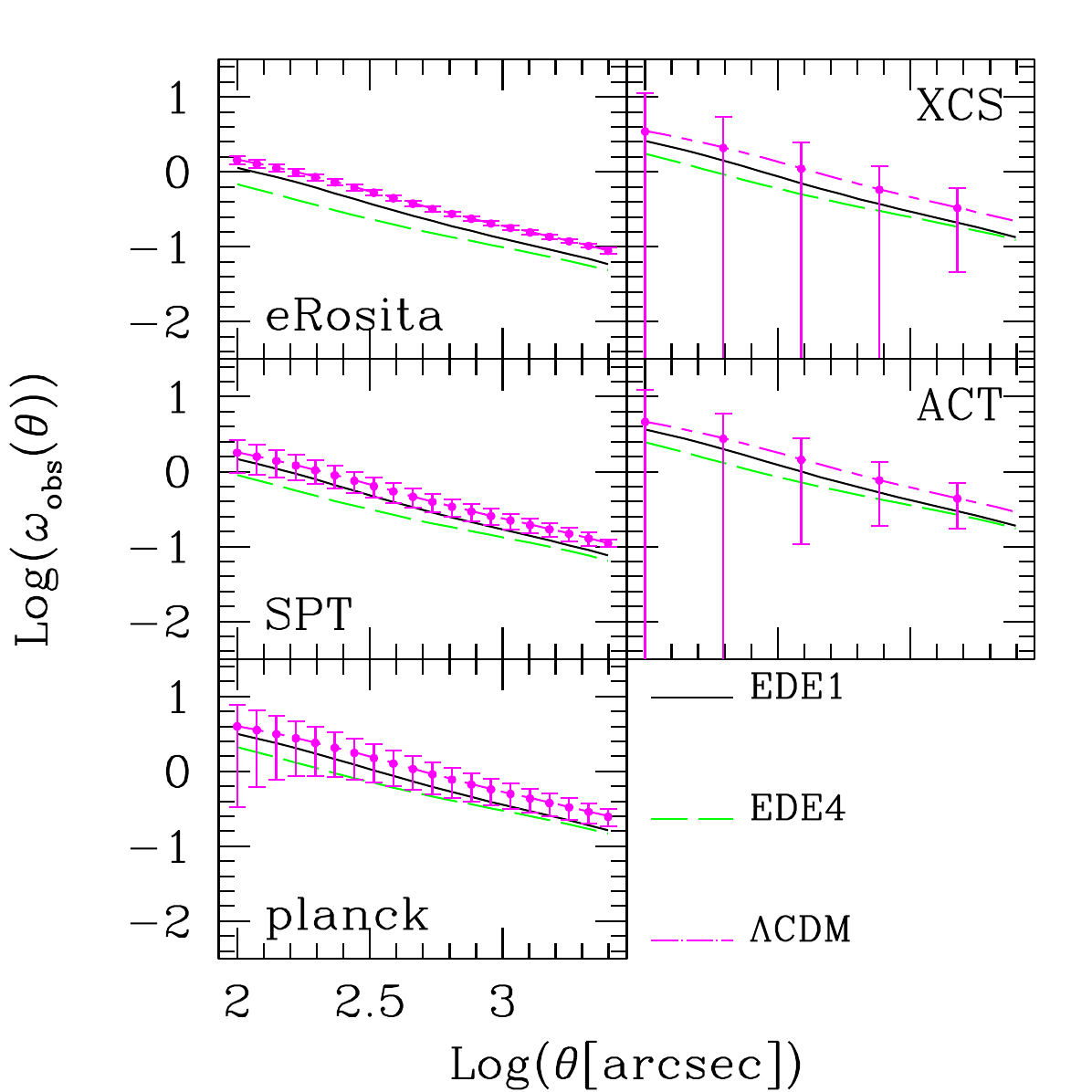}\hfill
\end{center}
\caption{As Fig.~\ref{fig:angular}, but with cluster catalogs restricted to $z \le 0.1$ (left panel) and $0.1< z \le 0.3$ (right panel). Only three cosmological models are shown here for simplicity.}
\label{fig:angular_z_lt}
\end{figure*}

\begin{figure}[t]
\begin{center}
  \includegraphics[width=\hsize]{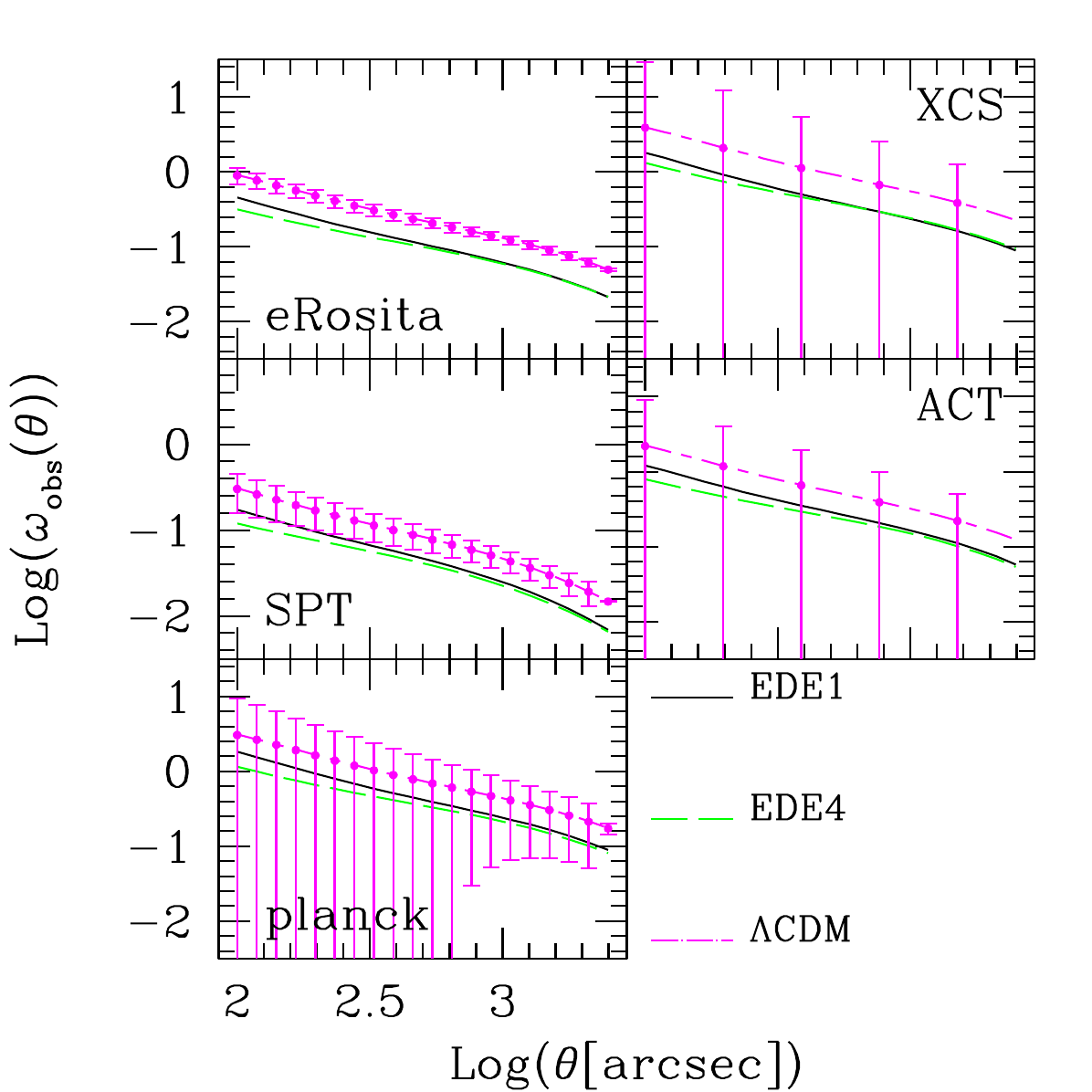}\hfill
\end{center}
\caption{As Fig.~\ref{fig:angular}, but only for catalogs restricted to $z > 0.3$. Only three cosmologies are shown here for simplicity.}
\label{fig:angular_z_gt}
\end{figure}

Figures~\ref{fig:angular_z_lt} and \ref{fig:angular_z_gt} show the angular correlation functions for cluster catalogs binned in redshift according to the scheme $z \le 0.1$, $0.1 < z \le 0.3$ and $z > 0.3$, as discussed above. For convenience, we show results for all the catalogs considered here and for the cosmological models EDE1, EDE4 and $\Lambda$CDM. In contrast to the spatial correlation functions, the angular correlation functions decrease in amplitude with increasing redshift, a trend that is significant  for the \emph{eRosita} and SPT samples, while the large relative errors for the \emph{Planck} catalog probably allow a detection of this decrement only at large angular separations and if the radial binning for the computation of error bars is enlarged.

The ratio of the correlation amplitudes in the $\Lambda$CDM model to those in the EDE models is slightly lower in the first two (low-redshift) bins than for the full samples. Since the relative error bars also increase in those bins, the signal-to-noise ratio for differences in the angular correlation functions is also lower. In contrast, the ratio increases slightly in the highest-redshift bin, $z > 0.3$, but there 
the relative errors are also larger, especially for \emph{Planck}. As a consequence, differences between the concordance model and both EDE models can be significantly detected with the \emph{eRosita} and, at least at large angular separations, with the SPT full sample or selected in the intermediate redshift bin, $0.1 < z < 0.3$. The situation remains again unchanged for XCS and ACT, with relative errors being too large to reach any significant conclusion.

According to the discussion above, distinguishing between different cosmologies by means of the angular correlation function does not take any particular advantage of redshift selection. It is legitimate to ask if the situation is different when selection is performed based on the observable used for cluster survey, instead of pure redshift. For instance, binning X-ray selected clusters according to their flux would correspond to a simultaneous (and highly non-linear) binning with mass and redshift. To check this issue, we performed clustering analysis of the SPT cluster sample after binning it according to flux density as $5$ mJy $\le S_{\nu_0} < 10$ mJy, $10$ mJy $< S_{\nu_0} \le 25$ mJy and $S_{\nu_0} > 25$ mJy. This choice ensures a suitably high number of spatial and angular pairs in each bin. However, the outcome is that the error bars are always larger than the differences between the various models for all bins. In other words, this kind of binning does not improve the detection of deviations from the concordance $\Lambda$CDM cosmological model with respect to the angular correlation function, and hence we decided not to show these results here.

\section{Summary and discussion}\label{sct:sum}

We have studied the clustering properties of galaxy clusters in various cosmological models with dynamical and non-dynamical evolution in the dark-energy component. In addition to the concordance $\Lambda$CDM cosmology, we addressed a model with a constant equation-of-state parameter for dark energy, $w_\mathrm{x} = -0.8$, a dynamical-dark energy model with the $w_\mathrm{x}(z)$ parametrization proposed by \cite{KO08.1}, and four models with a non-negligible amount of early dark energy. Cosmological models with dynamical dark energy are generically expected to form structures earlier, affecting both the abundance of massive galaxy clusters and their spatial distribution and clustering properties, this latter property having been the issue here.

To predict forthcoming observations, we computed the effective bias as a function of redshift and observed spatial and angular correlation functions as a function of (physical or apparent) separation that are expected to be measured in cluster catalogs produced by planned blind surveys both in the X-ray and in the sub-mm regimes by means of the thermal SZ effect. The X-ray surveys that we considered in this work are the \emph{eRosita} wide survey, which is described in detail in the related mission definition document, and the XMM cluster survey, based on existing pointings of the XMM satellite. For the SZ surveys, we focused on the South Pole Telescope, the Atacama Cosmology Telescope, and the \emph{Planck} all-sky survey of the Cosmic Microwave Background.

For computing the clustering properties of objects contained in each catalog as a function of cosmology, we employed a well-established formalism that takes past-light cone and selection effects on the cluster sample into account. To link the limiting flux of each survey to the minimum mass that enters the respective catalog at a given redshift, we adopted realistic scaling relations, based both on observations and numerical simulations, between the mass of a cluster and its X-ray luminosity or SZ flux density, or its integrated Compton-$y$ parameter. It turns out that the minimum mass entering a cluster catalog depends not only on the instrument considered, but also on the cosmological model adopted. This is so because the scaling relations mentioned above depend on the underlying world model through the expansion rate and cosmological distances.

As one could naively expect, the number of objects entering a given catalog at a given redshift depends heavily on the cosmology, the highest cluster abundances being present in models with EDE. This is caused by the well known higher mass function displayed by these models and the lower minimum mass entering the catalogs. The SPT catalog is the most extended in redshift due to the low limiting SZ flux density ($5$ mJy), while distributions for XCS and \emph{Planck} are the most limited in redshift, because of the quite shallow limiting X-ray flux of the former and large beam of the latter, which tends to dilute the signal.

For all catalogs, the first two models with early quintessence, EDE1 and EDE2, display the smallest effective bias at all redshifts. This is due to the high abundance of structures present in these models. On the other hand, the concordance $\Lambda$CDM model always displays the highest effective bias, for all catalogs and at all redshifts, while the other models lie somewhere in-between the two. Because of the width of their extent in redshift, SPT and ACT catalogs show the flattest trend for the effective bias, especially SPT, for which $b_\mathrm{eff}(z)$ is at most $\sim 10$ at $z = 3$. Other catalogs, particularly those based upon \emph{Planck} and XCS have a much steeper trend, the effective bias reaching considerably higher values already at $z \sim 2$.

Concerning the spatial correlation function, all the EDE models almost coincide on all scales, having a correlation amplitude smaller than that for the models $\Lambda$CDM, K08 and constant $w_\mathrm{x} = -0.8$, which are also very similar to each other. The largest difference is displayed by the catalogs produced with \emph{eRosita} and SPT. In the former, the relative errors are also smallest, thus making \emph{eRosita} the most promising instrument for the detection of EDE through the spatial correlation function, if the full cluster catalog is to be used. Detection might also be possible for SPT and \emph{Planck} at intermediate scales $r \sim r_0$, while it is completely out of question for XCS and ACT, whose error bars are too large. This is due to the shallow limiting fluxes for both of them, which is not adequately compensated for by the area covered (as is the case for \emph{Planck}). The situation is similar for the angular correlation function.  There the smallest error bars are produced by \emph{eRosita} as well, while the EDE4 model produces the smallest correlation amplitude for all the catalogs.

To explore the redshift evolution in the spatial and angular correlation functions, we also performed different cuts in redshift of the various cluster catalogs, in a way that the number of pairs of objects in each bin would remain approximately the same. For the number of spatial pairs, we adopted the double binning $z \le 0.1$ and $z > 0.1$, with the exception of SPT which allows the inclusion of a third bin, $z > 0.3$, with the modification of the second bin to $0.1 < z \le 0.3$. This same binning was also employed for the number of angular pairs.

The result of this analysis is that the spatial correlation function increases with increasing redshift, the correlation length for \emph{eRosita} growing by $\sim 80\%$ between the low- and high-redshift bins. The relative errors show that the increment in the correlation function is significant for \emph{eRosita}, SPT and \emph{Planck}, for all cosmological models considered in this work. Also, comparing the amplitude of the ratio between spatial correlation functions in EDE models and $\Lambda$CDM-like cosmologies, with the size of the error bars indicates that it is better to focus on the low-redshift cluster subsample in order to maximize significant differences between the concordance model and the EDE models in the \emph{eRosita} and \emph{Planck} catalogs. The high-redshift bin, $z > 0.3$, is better for the SPT catalog, giving results compatible with those from the full catalog. As for the angular correlation function, it tends to decrease with increasing redshift, a trend that is significant for all models and catalogs except the usual XCS and ACT. In the absence of errors, in the high-redshift bin there would be the highest chance of distinguishing a $\Lambda$CDM model from models with an early quintessence contribution, since there the ratios of correlation amplitudes are at their highest. However, the errorbars are also large there, so that \emph{eRosita} is the only survey expected to permit significant detection of deviations from the concordance model at all redshifts. In general, if one is interested in optimizing the differences between angular correlation functions measured in different models, it does not pay to subdivide catalogs according to redshift.

We also found that the same is true when cluster catalogs are binned according to the observable used to define them, an approach more directly motivated from the observational point of view. Specifically, binning the SPT catalog according to the flux density always produces ratios of angular correlation functions in different cosmologies that are similar to those for the full catalog, while the error bars are always larger.

Before concluding, two notes of caution are in order. First of all, some discussion arose in the literature on whether the semi-analytic calculations performed in \cite{BA06.1} on the spherical collapse model in cosmologies with dynamical-dark energy are indeed correct, and how accurately they are reproduced by $N$-body simulations \citep{FR08.1,FR08.2,GR08.1}. While the discussion is not yet settled, \cite{SA07.1} performed the same calculations as \cite{BA06.1} using a different approach, and found results that are perfectly consistent with the latter. In addition, \cite{SC08.1} used the same approach as \cite{BA06.1} to successfully evaluate the spherical collapse behavior in a modified gravity scenario. Hence, we are at least reasonably confident that the approach followed by \cite{BA06.1}, also employed in this work, is fundamentally correct. Moreover, numerical simulations using the correct early-time behavior of the growth factor for scaling the initial conditions yield results other than those of \cite{GR08.1} and \cite{FR08.1}, which tend towards the expectation from the analytic work of \cite{BA06.1}. Although precise direct integrations of the spherical collapse equations are difficult, further work avoiding approximations has so far confirmed our earlier results (Pace et al., in preparation). Even though definitive conclusions are not reached yet, these facts seem to justify our choice.

Secondly, estimates of the number counts of galaxy clusters detected by \emph{Planck} seem to fall substantially below the estimates of \cite{SC07.1}, and the redshift distribution is apparently shallower than that represented in Fig.~\ref{fig:dist} (\emph{Planck} SZ Challenge (in preparation), see also \citealt{LE08.1}). While definitive new limits for \emph{Planck} are not yet available, this part of the results should be read with caution. In particular, a decrease in the number of objects in the \emph{Planck} catalog would produce an increase in the relative errors, which scale as the inverse of the square root of the number of pairs of objects.

The present work shows that, while mild modifications to the redshift evolution in the equation-of-state parameter for dark energy $w_\mathrm{x}(z)$ have a negligible impact on the clustering properties of galaxy clusters, more exotic models such as early dark-energy cosmologies can change the effective bias of collapsed objects significantly, and thus also the spatial and angular correlation amplitudes of galaxy clusters. We have shown that at least some of the forthcoming blind surveys both in the X-ray and sub-mm regimes will be able to distinguish significantly between these models and more generally place constraints on the time evolution in the dark-energy density. 

In general, we expect object clustering to be less effective than e.g.,\ direct abundance data in determining the cosmological model. For instance, in the \emph{eRosita} catalog the number of clusters at $z \gtrsim 1$ increases by about one order of magnitude between the $\Lambda$CDM and the EDE4 models. This variation is much larger than the corresponding variation in the correlation functions, while the size of the relative errors, assuming Poisson statistics, is comparable. Object counting at high-redshift is also expected to be capable of distinguishing cosmological models with a gentle variation in dark-energy density from standard cosmology. The number of high-$z$ clusters in the \emph{eRosita} catalog is higher by a factor $\sim 2.5$ in the model with constant $w_\mathrm{x} = -0.8$ than for $\Lambda$CDM.

Nevertheless, the results of this work show that clustering of massive clusters by itself remains a fundamental channel to unravel the effect of the expansion history of the Universe on the process of structure formation, although employing this information in addition to simple object number counts is certainly an interesting issue to be explored.

{\small 
\section*{Acknowledgments}

We acknowledge financial contributions from contracts ASI-INAF I/023/05/0 and ASI-INAF I/088/06/0. We wish to thank the anonymous referee for useful remarks that allowed us to improve the presentation of our work.

\bibliographystyle{aa}
\bibliography{./master}
}

\end{document}